\documentclass[12pt]{jnmp}
\usepackage{amssymb, amsmath, xspace , 
}
\usepackage{epsfig}
\usepackage{amsmath}

\newcommand{\nad}[2]{\genfrac{}{}{0pt}{}{#1}{#2}}

\newcommand{\lb}[1]{\label{#1}}

\newcommand{\mref}[1]{(\ref{#1})}

\newcommand{\vf}{\varphi}

\renewcommand{\a}{\alpha}
\renewcommand{\b}{\beta}
\newcommand{\g}{\gamma}

\newcommand{\R}{\mathbb R}

\newcommand{\C}{\mathbb C}
\newcommand{\N}{\mathbb N}
\newcommand{\Z}{\mathbb Z}

\newcommand{\cdva}{{{\cal C}_{2}(m,l)}}
\newcommand{\cnm}{{{\cal C}_{n}(l,m)}}

\newcommand{\dtilde}{\raisebox{3.5mm}{$\approx$}\hskip-3.5mm}
\newcommand{\tips}{{T_i^+}^s}
\newcommand{\tims}{{T_i^-}^s}
\newcommand{\ajp}{a_j^+}
\newcommand{\tjp}{T_j^+}
\newcommand{\ajm}{a_j^-}
\newcommand{\tjm}{T_j^-}
\newcommand{\timspo}{{T_i^-}^{s+1}}
\newcommand{\aim}{a_i^-}
\newcommand{\aip}{a_i^+}
\newcommand{\ajpm}{a_j^\pm}
\newcommand{\aipm}{a_i^\pm}
\newcommand{\tjps}{{T_j^+}^s}
\newcommand{\tip}{T_i^+}
\newcommand{\aqpm}{a_q^\pm}
\newcommand{\aqp}{a_q^+}
\newcommand{\tqp}{T_q^+}
\newcommand{\aqm}{a_q^-}
\newcommand{\tqm}{T_q^-}
\newcommand{\tipspo}{{T_i^+}^{s+1}}
\newcommand{\tjpspo}{{T_j^+}^{s+1}}
\newcommand{\tipsmo}{{T_i^+}^{s-1}}
\newcommand{\timsmo}{{T_i^-}^{s-1}}
\newcommand{\tim}{T_i^-}
\newcommand{\tjpsmo}{{T_j^+}^{s-1}}
\newcommand{\tnp}{T_n^+}
\newcommand{\anp}{a_n^+}
\newcommand{\anm}{a_n^-}
\newcommand{\tnm}{T_n^-}

\renewcommand{\le}{\leqslant}
\renewcommand{\ge}{\geqslant}

\newtheorem{predl}{Proposition}
\newtheorem{theorem}{Theorem}
\newtheorem{lemma}{Lemma}

\theoremstyle{definition}
\newtheorem{definition}{Definition}
\newtheorem{remark}{Remark}

\def\beq#1#2\eeq{%
        \begin{equation}%
        \label{#1}%
            #2%
        \end{equation}%
    }

\JNMPnumberwithin{equation}{section}
\begin{document}

\renewcommand{\evenhead}{M Feigin}
\renewcommand{\oddhead}{Bispectrality for deformed CMS systems}

\Name{Bispectrality for deformed  Calogero--Moser--Sutherland systems }

\Author{Misha Feigin}

\Address{Department of Mathematics, South Kensington campus, Imperial College London, London SW7 2AZ, UK \\
E-mail: m.feigin@imperial.ac.uk}

\begin{abstract}
\noindent We prove bispectral duality for the generalized
Calogero--Moser--Sutherland systems related to configurations
${\cal A}_{n,2}(m), \ {\cal C}_n(l,m)$. The trigonometric axiomatics of
Baker--Akhiezer function is modified, the dual difference
operators of rational Macdonald type  and the Baker--Akhiezer
functions related to both series are explicitly constructed.
\end{abstract}

\section{Introduction}

Original bispectral problem as it appeared in the paper by
Duistermaat and Grunbaum \cite{DG} was devoted to investigations
of such  Sturm--Liouville operators that admit a family of
eigenfunctions satisfying some differential equation in the
spectral parameter. Part of the corresponding potentials, namely
the rational KdV potentials, were described as those which can be
obtained from 0 by applying Darboux transformations. The
corresponding Sturm--Liouville operators admit non-trivial
commuting differential operators. In paper \cite{DG} the
conditions in terms of local Loran expansions for a potential to
be a rational KdV potentials were also analyzed. These conditions
generalized simple locus conditions from \cite{AMM}.

An example which may be looked at as generalization of this
picture to many-dimensional case is given by Calogero--Moser
operator (\cite{C}, \cite{M}, \cite{S}, \cite{OP}) \beq{cm} L= \Delta -
\sum_{\a \in A} \frac{m_\a (m_\a+1)(\a,\a)}{(\a,x)^2}, \eeq where
$A$ is a root system. When the parameters $m_\a$ are integer (and
invariant), as it was discovered by Chalykh and Veselov
\cite{ChV1}, operator \mref{cm} can be included into a large
supercomplete commutative ring of differential operators. The key
object of this construction is the multidimensional
Baker--Akhiezer function
$\psi(k,x)=\psi(k_1,\ldots,k_n,x_1,\ldots,x_n)$. This function is defined
on certain many-dimensional rational spectral variety, and it is an
eigenfunction for all operators from the commutative ring. Function $\psi$
satisfies same differential equations in the variables
$k$, as it was shown by Chalykh, Styrkas, and Veselov \cite{VSCh}, thus bispectrality holds. Baker--Akhiezer functions defined on riemannian
surfaces were introduced by Krichever for studying one variable rings of commuting differential operators and non-linear integrable equations \cite{K}
(see also \cite{BC}).

Generalization of the one-dimensional locus conditions from
\cite{DG} to multi-dimensional case led Chalykh, Veselov, and the
author to an interesting class of Schroedinger operators
containing operators \mref{cm} where $A$ can be a Coxeter system
and also more general locus configuration \cite{CFV3}. These
operators can be included into supercomplete rings of commuting
differential operators, they admit Baker--Akhiezer functions which
also satisfy differential equations in the spectral parameters.

Despite large number of locus configurations in the
two-dimensional case the known examples in higher dimensions are
quite exceptional and discreet. Besides operators \mref{cm}
related to Coxeter systems two other series of deformations
$A_{n,1}(m)$, $\cnm$ were found in \cite{CFV1},
\cite{CFV2},\cite{CFV3}, and one more configuration $A_{n,2}(m)$
appeared later in Chalykh--Veselov paper \cite{CV2}. Configuration $A_{n,1} (m)$ becomes
root system $A_n$ when m=1. Configuration $\cnm$ specializes to
root system $C_n$ at $l=m$. Configuration $A_{n,2} (m)$ is a
complex extension of the root system $A_{n-2}$. When $n=2$
parameter $m$ can be arbitrary complex and the corresponding
operator coincides with the degeneration of the Hietarinta
operator \cite{H} (see also \cite{CFV3}, \cite{CEO}). In this
paper we analyze bispectrality and Baker--Akhiezer functions for the
trigonometric versions of the operators \mref{cm} for the
configurations $\cnm$ and $A_{n,2}(m)$, whereas root systems and
configuration $A_{n,1}(m)$ were considered by Chalykh
\cite{ChalBisp}. Earlier in paper \cite{CFV2} the intertwining
operators for the Schroedinger operators with trigonometric
potentials related to $\cdva$, $A_{n,1}(m)$  were constructed.

Construction with multi-dimensional trigonometric Baker--Akhiezer
function was also introduced by Chalykh and Veselov in \cite{ChV1}.
Such a function is certain eigenfunction for the generalized
Calogero--Moser--Sutherland  operator
\beq{cmtr} L= \Delta -
\sum_{\a \in A} \frac{m_\a (m_\a+1)(\a,\a)}{\sinh^2(\a,x)}. \eeq
It was shown in \cite{ChV1} that Baker--Akhiezer function exists
when $A$ is a root system and multiplicities $m_\a$ are integer
and invariant. Then $L$ is included into a supercomplete ring of
commuting differential operators.

In the trigonometric case the dual operators happen to be
difference operators. These operators are also discretization of
Calogero--Moser hamiltonians, they were introduced by Ruijsenaars
for the problem related to $A_n$ root system \cite{R1} (see \cite{RS} for
the classical version). The bispectral duality of
Calogero--Moser--Sutherland and Ruijsenaars systems was
conjectured by Ruijsenaars in \cite{R2}. For arbitrary reduced
root system the difference operators were introduced by Macdonald
\cite{Macdonald}. The duality on the level of Macdonald polynomials was
conjectured by Macdonald and proved first by Koornwinder
\cite{Koorn1} (see chapter VI of \cite{Mac})
for $A_n$ case. For arbitrary reduced root systems the proof was
obtained by Cherednik \cite{Cher}. For the case of $BC_n$ system
Macdonald polynomials were introduced by Koornwinder
\cite{Koorn2}, their duality property was established in
\cite{vD}, \cite{Sahi}.

In terms of Baker--Akhiezer functions the bispectral duality for
\mref{cmtr} related to any root system was established by Chalykh
in \cite{ChalBisp}. Also it was done for the system $A_{n,1}$ thus
the corresponding deformation of rational Ruijsenaars--Macdonald
operator appeared in \cite{ChalBisp}.

The method of establishing dual equations as well as of
constructing the Baker--Akhiezer functions was introduced by
Chalykh in \cite{ChalBisp}, and it is as follows. The
Baker--Akhiezer function should satisfy some shifting conditions
as a function of spectral variables $k$. One considers space of
functions satisfying these conditions and some difference operator
in $k$ such that the application of this operator leaves the space
invariant. Then taking a proper initial function from the space
and iterating the application of operator we arrive to the
Baker--Akhiezer function, besides that on the next step we get
zero thus the dual equation appears. This method was first applied
in the rational case \cite{Ch} (see also \cite{CFV3}), it works
also in the trigonometric difference case \cite{Ch3}. The
corresponding formula for the Baker--Akhiezer functions in the rational case
was found earlier by Berest \cite{B} under assumption of existence.

In this paper we follow the described strategy to construct
Baker--Akhiezer functions and establish the bispectrality for the
configurations $\cnm$, $A_{n,2}(m)$. On the way we introduce the
generailzations of rational Macdonald operators related to these
deformations. An interesting feature of configuration $A_{n,2}(m)$
is that for the corresponding operator \mref{cmtr} there is no
Baker--Akhiezer function in the original axiomatics \cite{ChV1}.
Thus we modify conditions in variables $k$ which should be imposed on the
Baker--Akhiezer function in order to cover this case as well. The
corresponding modification of rational Chalykh--Veselov axiomatics
for Baker--Akhiezer functions from \cite{ChV1} was carried out in
\cite{CFV3}. We should mention that in our considerations we
restrict ourselves to the simpler case when configuration $A$ has
no parallel vectors although a deformation of $BC_n$ system leading to
algebraically integrable operators appeared in \cite{CEO}, so it
is natural to expect the bispectral property for the degeneration of this
 model as well.

The structure of this paper is the following. In section 2 we give
modified axiomatics for trigonometric Baker--Akhiezer function and
review Chalykh--Veselov construction \cite{ChV1} adopting it to
the new settings. In section 3 we recall how the bispectrality
allows construction of commuting operators in the spectral
variables if we know commuting operators in $x$ (\cite{B}, \cite{DG},  \cite{ChalBisp}).
 Then we prove
that the Baker--Akhiezer function for root systems and deformation
$A_{n,1}(m)$ also satisfy modified axiomatics. In section 4 we
consider configuration $\cnm$. We introduce deformed rational
Macdonald operator for this case and we explicitly construct
Baker--Akhiezer function. Then we prove bispectral property, and
the family of commuting difference operators appears. In section 5
the analogous results are proved for the $A_{n,2}(m)$
configuration. In the last section we discuss necessary conditions
for a configuration of vectors with multiplicities to admit
Baker--Akhiezer function. They reveal clear geometrical restrictions on the
configurations. The presentation closely follows
\cite{Phd}.

\vspace{5mm} {\bf Acknowledgements.} I am very grateful to O.A.
Chalykh for numerous stimulating and helpful discussions, and to
A.P. Veselov for useful discussions and interest to the work. On
the final stage the work was supported by Chapman Fellowship (Department
of Mathematics, Imperial College) and by the European grant ``ENIGMA" (MRTN-CT-2004-5652).

\section{Baker--Akhiezer function and commuting differential operators}
\label{gl2sect1}

Let  $A$ be a finite set  of non-collinear vectors $\alpha\in
\mathbb C^n$, let every vector $\a$ have multiplicity $m_\alpha
\in \mathbb N$. Meaning by $m$ this multiplicity function we will
denote such configurations as ${\cal A} = (A, m)$. By {\it
Baker--Akhiezer function} $\psi(k,x)$ we will mean a function of
two sets of variables \ $k,x\in\mathbb C^n$ of the form
\begin{equation}\label{psi12}
\psi(k,x)=\Bigl(\prod_{\alpha\in A} (k,\alpha)^{m_\alpha}+\mbox{lower order
polynomial in} \, k \Bigr)
e^{(k,x)},
\end{equation}
which satisfies special properties. We introduce $-A$ to be the system of vectors
$\{-\alpha| \alpha\in
A\}$ with the multiplicities $m_{-\alpha}=m_\alpha$. Inside $A\cup -A$ we
 choose a positive subsystem $A_+$ consisting of those vectors which belong to some
 half-space inside $\mathbb R^{2n}\approx \mathbb C^n$. The half-space should
 be  in generic position such that for any $\alpha\in A$ either $\alpha\in A_+$  or $-\alpha\in A_+$. We say that a vector  $\alpha\in A_+$ ÿis {\it an  edge  vector } if $\a$ is not a linear combination of other vectors from $A_+$ with  positive real coefficients.

In this paper we will assume that the set $A$ of vectors $\a$ is such that
all the vectors belong to some lattice of rank $n$ in the space $\C^n$. Though
constructions and most of the proofs work without this assumption in all known examples such a lattice does exist, also assumption on the lattice makes definition of the edge vectors and subsystems $A_+$ more invariant. Namely, we now have an $n$-dimensional real vector space $V$ containing the system $A$ which is spanned by a basis in the lattice. Positive subsystems $A_+ \subset (A\cup -A)$ are those which consist of vectors belonging to a generic half-space in the real linear space $V$. We will also assume that
$A$ does not contain isotropic vectors $\a: (\a,\a)=0$, as we will see such
vectors do not contribute to the potential.

\begin{definition}
{A function $\psi(k,x)$ of the form $(\ref{psi12})$ is called the
{\it Baker--Akhiezer function } for a configuration
 ${\cal A} = (A, m)$ (BA function) if for any choice of positive subsystem
 $A_+$ and for any choice of an edge vector $\alpha$ the following identity
 holds
\begin{equation}\label{22}
\frac{\psi(k+s\alpha,x)}{\prod_{\nad{\beta\in A_+}{\beta\ne \alpha}}
\prod_{i=1}^{m_\beta} (k+i\beta+s\alpha,\beta)}
\equiv
\frac{\psi(k-s\alpha,x)}{\prod_{\nad{\beta\in A_+}{\beta\ne \alpha}}
\prod_{i=1}^{m_\beta} (k+i\beta-s\alpha,\beta)}
\end{equation}
at $(k,\alpha)=0$, $s=1,\dots,m_\alpha$.}
\end{definition}
\begin{remark}
For a given vector $\a\in A$ there are normally few choices of the subsystems
$A_+$ such that the vector $\a$ is an edge vector. Therefore existence of
Baker--Akhiezer function for a system $A$ forces, in particular,  the following compatibility conditions. Let $A_+^1, A_+^2$ be two choices
of positive subsystems in $A$ such that $\a$ is an edge vector. Then the
following identity must hold:
$$
\frac{
\prod_{\nad{\beta\in A_+^1}{\beta\ne \alpha}}
\prod_{i=1}^{m_\beta} (k+i\beta+s\alpha,\beta)
}
{
\prod_{\nad{\beta\in A_+^1}{\beta\ne \alpha}}
\prod_{i=1}^{m_\beta} (k+i\beta-s\alpha,\beta)
}
\equiv
\frac{
\prod_{\nad{\beta\in A_+^2}{\beta\ne \alpha}}
\prod_{i=1}^{m_\beta} (k+i\beta+s\alpha,\beta)
}
{
\prod_{\nad{\beta\in A_+^2}{\beta\ne \alpha}}
\prod_{i=1}^{m_\beta} (k+i\beta-s\alpha,\beta)
}
$$
at $(k,\alpha)=0$ for $s=1,\dots,m_\alpha$.
\end{remark}
Introducing  the functions $\psi_\alpha^{A_+}$ depending on the choices of
positive subsystem $A_+$ and an edge vector $\a$ by formulas
$$
\psi_\alpha^{A_+}=
\frac{\psi(k,x)}{\prod_{\nad{\beta\in A_+}{\beta\ne \alpha}}
\prod_{i=1}^{m_\beta} (k+i\beta,\beta)}
$$
conditions~(\ref{22}) take the following form
\begin{equation}\label{2'}
\psi_\alpha^{A_+}(k+s\alpha)\equiv \psi_\alpha^{A_+}(k-s\alpha), \qquad
\mbox{if}\quad  (\alpha,k)\equiv 0,\  s=1,\dots,m_\alpha.
\tag {$\ref{22}'$}
\end{equation}
Also it will be convenient for us to use the following equivalent form of
equations~(\ref{22})
\begin{equation}\label{2''}
\biggl(\delta_\alpha
\frac{1}{(k,\alpha)}\biggr)^{s-1}\delta_\alpha\psi_\alpha^{A_+}\equiv 0,
\qquad
\mbox{ at }  (k,\alpha)=0,\  s=1,\dots,m_\alpha.
\tag {$\ref{22}''$}
\end{equation}
Here $\delta_\alpha$ is an operator acting by the rule  $\delta_\alpha f(k)=f(k+\alpha)-f(k-\alpha)$.
It is obvious that conditions~(\ref{2'}) and~(\ref{2''}) are identical for
$m_\a=1$. One can also check that they are equivalent in general.
Conditions
~(\ref{22}) form an overdetermined system of equations for the coefficients
of a polynomial in~(\ref{psi12}). It takes place the following statement.

\begin{predl} $(c.f. \cite{ChV1})$
If the Baker--Akhiezer function exists then it is unique.
\end{predl}

\begin{proof}
Assume there are two functions
$\varphi_1=P_1(k,x)  e^{(k,x)}$,
$\varphi_2=P_2(k,x) e^{(k,x)}$,
which satisfy equations
(\ref{22}), and assume the  highest terms of the polynomials $P_i(k,x)$ are
$\prod_{\alpha\in A} (k,\alpha)^{m_\alpha}$. Consider the difference
$\varphi_1-\varphi_2=(P_1-P_2)e^{(k,x)}$. This function also satisfies conditions
(\ref{22}) but the degree of polynomial $P_1-P_2$ is less than $\sum_{\alpha\in A} m_\alpha$. The proof of the proposition thus reduces to the following
statement.

\begin{lemma}\label{lem1} $(c.f.\cite{ChV1})$
Let $\psi(k,x)=P(k,x) e^{(k,x)}$ satisfy conditions $(\ref{22})$ with
$P(k,x)$ being a polynomial in $k$ with the highest term $P_0(k,x)$. Then $P_0(k,x)$
is divisible by $\prod_{\alpha\in A} (k,\alpha)^{m_\alpha}$.
\end{lemma}

\begin{proof} Consider condition~(\ref{2''}) for some subsystem $A_+$ and an edge
vector $\a$. We have
$$
\psi_\alpha^{A_+}=\frac{P(k,x)}{Q(k)} e^{(k,x)},
$$
where
$$
Q(k)=\prod_{\nad{\beta\in A_+}{\beta\ne\alpha}}\prod_{i=1}^{m_\beta} (k+i\beta,\beta).
$$
We denote by $Q_0(k)$ the highest term of $Q(k)$ and consider conditions
(\ref{2''}) with $s=1$
\begin{multline*}
\frac{P(k+\alpha,x)}{Q(k+\alpha)}\, e^{(\alpha, x)} e^{(k,x)}-
\frac{P(k-\alpha,x)}{Q(k-\alpha)}\, e^{-(\alpha, x)} e^{(k,x)}
=\\= e^{(k,x)} \frac{P_0(k,x)\,(e^{(\alpha, x)}-e^{(-\alpha,
x)})\, Q_0(k) + \mbox{ lower terms}}{Q(k+\alpha)Q(k-\alpha)} \quad
\vdots\  (k,\alpha).
\end{multline*}
As $Q_0(k)=\prod_{\nad{\beta\in
A_+}{\beta\ne\alpha}}(k,\beta)^{m_\beta}$ is not divisible by
$(k,\alpha)$ we conclude that $P_0(k,x)$ should be divisible by
$(k,\alpha)$.

Now let's rewrite the obtained relation in the form
$$
\delta_\alpha \psi_\alpha^{A_+}=
(k,\alpha)\frac{\widetilde P(k,x)}{\widetilde Q(k)}\, e^{(k,x)} (e^{(\a,x)}
- e^{(-\a,x)}),
$$
where $\widetilde P$, $\widetilde Q$ are some polynomials in $k$ with the highest terms $\widetilde P_0 = \frac{P_0}{(k,\a)}Q_0$, and $\widetilde Q_0=\prod_{\nad{\beta\in A_+}{\beta\ne\alpha}} (k,\b)^{2 m_\b}$,
so $\widetilde Q_0$ is again not divisible by  $(k,\a)$.
Considering conditions~(\ref{2''}) with $s=2$ we analogously
conclude that $\widetilde P_0\ \vdots\ (k,\alpha)$, that is $P_0\ \vdots\ (k,\alpha)^2$. Continuing in this way
up to $s=m_\alpha$ we obtain $P_0\ \vdots\ (k,\alpha)^{m_\alpha}$.
Since any vector $\alpha\in A$ is an edge vector for the proper
choice of subsystem
 $A_+$, system~(\ref{2''})  contains equations for all  $\alpha\in A$. Therefore
$P_0\ \vdots\ \prod_{\alpha\in A} (k,\alpha)^{m_\alpha}$, and lemma is proven.
\end{proof}
\end{proof}

The existence of Baker--Akhiezer function is possible for very special configurations
${\cal A}$ only.
In this case
 $\psi(k,x)$
becomes a joint eigenfunction of a rich commutative ring of differential
operators. Namely to any configuration  ${\cal A}$ let us relate ring $R_{\cal A}$ of polynomials $p(k)$ which for any $\alpha\in A$ satisfy conditions
$$
p(k+s\alpha)\equiv p(k-s\alpha) \quad \mbox{at} \,\, (k,\alpha)=0,
$$
where $s=1,\dots,m_\alpha$.

\begin{theorem} \lb{nt} $(c.f. \cite{ChV1})$
Assume configuration ${\cal A}$ admits Baker--Akhiezer function.
Then for any $p(k)\in R_{\cal A}$ there exists differential
operator $L_p(x,\p_x)$ such that
$$
L_p(x,\p_x) \, \psi (k,x)=p(k)\, \psi(k,x).
$$
And for any $p,q\in R_{\cal A}$ one has commutativity $L_p L_q=L_q L_p$.
\end{theorem}

\begin{proof} Consider function $\psi_1(k,x)=p(k)\, \psi(k,x)-p(\p_x)\psi(k,x)$.  Then function
$\psi_1$ satisfies conditions~(\ref{22})  and it has the form
$\psi_1=Q_1(k,x) e^{(k,x)}$ with $\deg Q_1\le \sum m_\alpha +\deg
p -1$. By lemma \ref{lem1} the highest term of polynomial $Q_1$ has the form
$Q_1^0=\prod (k,\alpha)^{m_\alpha} r(x,k)$. We define now
$\psi_2(k,x)=\psi_1(k,x)-r(x,\p/\p x)\,
\psi(k,x)$. We have $\psi_2(k,x)=Q_2(k,x)e^{(k,x)}$, where
$Q_2$ is some polynomial of degree
$\deg Q_2 \le \sum m_\alpha +\deg p-2$, and $\psi_2$ satisfies conditions~(\ref{22}).
Therefore we can again apply lemma  \ref{lem1} and inductively we construct
operator $L_p=p(\p_x)+r(x,\p_x)+\dots$

Commutativity $[L_p,L_q]=0$ follows from the condition that if an
operator $L(x,\p_x)$ satisfies condition $L(x,\p_x)\psi(k,x)=0$
for a function $\psi$ of the form~(\ref{psi12}), then $L\equiv 0$.
Theorem is proven.
\end{proof}

We note that for any configuration
${\cal A}$ the ring $R_{\cal A}$ contains polynomial
 $k^2=k_1^2+\ldots+k_n^2$. Indeed,
$(k\pm s\alpha)^2=(k\pm s\alpha,k\pm s\alpha)=(k,k)\pm
2s(\alpha,k)+s^2(\alpha,\alpha)$,
and if $(\alpha,k)=0$ we have $(k+s\alpha)^2=(k-s\alpha)^2$. The corresponding
differential operator is the Schroedinger operator.

\begin{predl}\label{urS} $(c.f. \cite{ChV1})$
In the settings of theorem \ref{nt} to polynomial $p(k)=k^2$ it corresponds operator
$$
L_{k^2}=\Delta-\sum_{\alpha\in A}
\frac{m_\alpha (m_\alpha+1) (\alpha,\alpha)}{\sinh^2 (\alpha,x)}.
$$
\end{predl}

\begin{proof}
Let
$$
\psi(k,x)= P(k,x) e^{(k,x)}=
(\prod_{\a\in A} (k,\alpha)^{m_\alpha}+P_1+ \mbox{lower order terms})e^{(k,x)},
$$
where $P_1$ is polynomial of degree $\sum m_\alpha -1$.
To obtain  $L_{k^2}$ we apply recurrent procedure described in the proof
of theorem \ref{nt}. We have
\begin{multline*}
\psi_1(k,x)=k^2\psi(k,x)-\Delta\psi(k,x)
=\\=
\biggl(-2\sum_{i=1}^n k_i\frac{\p}{\p x_i}\, P -\Delta P\biggr) e^{(k,x)}
=
\biggl(-2\sum_{i=1}^n k_i\frac{\p P_1}{\p x_i} +R \biggr) e^{(k,x)},
\end{multline*}
where $R$ is some polynomial in $k$, $\deg R<\sum m_\alpha$.
According to lemma~\ref{lem1}
$$
-2\sum_{i=1}^n k_i\frac{\p P_1}{\p x_i}= u(x) \prod_{\alpha\in A}(k,\alpha)^{m_\alpha}
$$
for some function $u(x)$. Also from lemma \ref{lem1} it follows
that $\psi_1(k,x)-u(x)\psi(k,x)=0$. Thus
$$
L_{k^2}=\Delta+u=\Delta-\frac{2}{\prod_{\alpha\in A}(k,\alpha)^{m_\alpha}}
\sum_{i=1}^n
k_i \frac{\p P_1}{\p x_i}.
$$
And the proof of the proposition is reduced to the following lemma

\begin{lemma}\lb{lmm2} $(c.f. \cite{ChV1})$
Assume that a system ${\cal A}$ admits Baker--Akhiezer function
$$
\psi(k,x)=P(k,x) e^{(k,x)}=
\Bigl(\prod_{\alpha\in A} (k,\alpha)^{m_\alpha}+P_1+\ldots\Bigr)e^{(k,x)},
$$
where $P_1=P_1(k,x)$ are terms of order  $\sum m_\alpha -1$ in the
polynomial $P$. Then
$$
P_1=-\bigg(\prod_{\alpha\in A}(k,\alpha)^{m_\alpha}\bigg) \sum_{\alpha\in A}
\frac{m_\alpha(m_\alpha+1)}{2}\,
\frac{(\alpha,\alpha)}{(\alpha,k)}\,
\coth (\alpha,x).
$$
\end{lemma}

\begin{proof}
Let us choose a subsystem $A_+$ and consider
conditions~(\ref{2''}) for arbitrary edge vector $\a$. We want to
show that $P_1$ is divisible by $(k,\alpha)^{m_\alpha-1}$  and to
find $P_1/(k,\alpha)^{m_\alpha-1}$. For $s=1$
condition~(\ref{2''}) can be rewritten in the following way
\begin{multline*}
\frac{1}{Q_1(k)}
\biggl\{
T_1(k)(k,\alpha)^{m_\alpha}\prod_{\nad{\beta\in A}{\beta\ne \alpha}} (k,\beta)^{m_\beta}
(e^{(\a, x)}-e^{-(\a, x)})
+ \\+
T_1(k)\biggl(
m_\alpha (\alpha,\alpha) (k,\alpha)^{m_\alpha-1}
(e^{(\a, x)}+e^{-(\a, x)})
\prod_{\nad{\beta\in A}{\beta\ne \alpha}}  (k,\beta)^{m_\beta} +
(e^{(\a, x)}-e^{-(\a, x)}) P_1\biggr)
+\\+
O_1\bigl((k,\alpha)^{m_\alpha}\bigr)+R_1
\biggr\}
\, e^{(k,x)}\equiv 0,
\end{multline*}
if $(k,\alpha)=0$. In the last formula
$$
T_1(k)=\prod_{\nad{\beta\in A_+}{\beta\ne \alpha}}  \prod_{i=1}^{m_\beta}
(k+i\beta,\beta),
\qquad
Q_1(k)=T_1(k+\alpha)\,T_1(k-\alpha),
$$
and
$ O_1\bigl((k,\alpha)^{m_\alpha}\bigr)$
is a polynomial
of degree $2\sum_{\beta\in A_+} m_\beta-1-m_\alpha$, which is divisible by
$(k,\alpha)^{m_\alpha}$. And $R_1$ is some polynomial in $k$ such that
$\deg R_1<2\sum_{\beta\in A_+} m_\beta-1-m_\alpha$.

Going by induction we conclude that for arbitrary
 $s$ such that $m_\alpha\ge s>1$ one has
\begin{multline}\lb{nmt}
\frac{1}{Q_s(k)}
\biggl\{
T_s(k)(k,\alpha)^{m_\alpha-s+1}\prod_{\nad{\beta\in A}{\beta\ne \alpha}}  (k,\beta)^{m_\beta}
+\\+
T_s(k)\biggl(
c_s(m_\alpha) (\alpha,\alpha) (k,\alpha)^{m_\alpha-s}
\prod_{\nad{\beta\in A}{\beta\ne \alpha}}  (k,\beta)^{m_\beta} \coth(\alpha,x)
+\frac{P_1}{(k,\alpha)^{s-1}}\biggr)
+\\+
O_s\bigl((k,\alpha)^{m_\alpha-s+1}\bigr) +R_s
\biggr\}
\equiv0
\end{multline}
if $(k,\alpha)=0$. Here
$T_s(k)=T_{s-1}(k)\, Q_{s-1}(k)$,
$Q_s(k)=Q_{s-1}(k+\alpha)\, Q_{s-1}(k-\alpha)$,
and polynomial $O_s\bigl( (k,\alpha)^{m_\alpha-s+1}\bigr)$
is divisible by
$(k,\alpha)^{m_\alpha-s+1}$,
$\deg O_s\le \deg T_s+\sum_{\beta\in A_+}m_\beta-s$,
$\deg R_s < \deg T_s+\sum_{\beta\in A} m_\beta-s$. It is important for us
that
$c_s(m_\alpha)=c_{s-1}(m_\alpha)+m_\alpha-s+1$. Consider now condition (\ref{nmt}) with $s=m_\alpha$. As $T_s(k)\ne 0$ if $(k,\alpha)=0$ we conclude that
\begin{equation}\label{4}
c_{m_\alpha} (m_\alpha) (\alpha,\alpha)
\prod_{\nad{\beta \in A}{\beta\ne \alpha}} (k,\beta)^{m_\beta} \coth (\alpha,x)
+ \frac{P_1}{(k,\alpha)^{m_\alpha-1}}=0
\end{equation}
at $(\alpha,k)=0$, and
$$
c_{m_\alpha}(m_\alpha)=m_\alpha+(m_\alpha-1)+\ldots+1=
\frac{m_\alpha(m_\alpha+1)}{2}.
$$
Now we remark that conditions~(\ref{4}) characterize polynomial
$P_1$ uniquely. Indeed the existence of a polynomial $\widetilde
P_1$, $\deg \widetilde P_1=\deg P_1=\sum m_\beta -1$,
satisfying~(\ref{4}) would mean that $\frac{P_1-\widetilde
P_1}{(k,\alpha)^{m_\alpha-1}}=0$ at $(k,\alpha)=0$, thus
$P_1-\widetilde P_1$ would be divisible by
$(k,\alpha)^{m_\alpha}$. As any vector $\alpha\in A$ is an edge vector
for a proper subsystem~${A_+}$, we get $P_1-\widetilde P_1\
\vdots\ \prod_{\alpha\in A} (k,\alpha)^{m_\alpha}$. But this is
impossible as \hbox{$\deg (P_1-\widetilde P_1)\le \sum_{\alpha\in
A}m_\alpha-1$}. Further it is obvious that polynomial
$$
P_1=-\bigg(\prod_{\alpha\in A}(k,\alpha)^{m_\alpha}\bigg)
\sum_{\alpha\in A} \frac{m_\alpha (m_\alpha+1)}{2}\, \frac{(\alpha,\alpha)}{(\alpha,k)}
\coth (\alpha,x)
$$
satisfies~(\ref{4}), therefore lemma \ref{lmm2} is proven.
\end{proof}
And this completes proof of proposition.
\end{proof}

\section{Bispectral duality and examples}
\label{gl2sect2}

By bispectral duality we mean the situation when a function
$\psi(k,x)$ of two sets of variables $k$ and $x$ satisfies certain
equations in each of the sets. In our case we'll have the
equations of the form
\begin{equation}
\label{bisp} L(x,\p_x) \psi (k,x) = k^2 \psi(k,x),\\
D \psi (k,x) = \lambda(x) \psi(k,x),
\end{equation}
where $D$ is some {\it difference} operator in $k$ variables, and
$\psi$ is the Baker-Akhiezer function. Originally the equations in
the spectral parameter were considered by Duistermaat and Grunbaum
\cite{DG} who analyzed in the one-dimensional situation the pair
of equations \mref{bisp} for Sturm--Liouville operator $L$ and
differential operator $D$.

One of the applications of the bispectrality is the following
construction  (\cite{DG}, \cite{B}, \cite{ChalBisp}) allowing to
obtain commuting operator for $D$ if a commuting operator for $L$
is given. More exactly, assume we have some operator $M(x, \p_x)$
satisfying
\begin{equation}
\label{bispM} M(x,\p_x) \psi (k,x) = q(k) \psi(k,x),
\end{equation}
for some polynomial $q(k)$.
Then from  \mref{bisp}, \mref{bispM}  it follows
$$
(\lambda M - M \lambda) \psi (k,x) = (q D - D q) \psi (k,x)
$$
Iterating this process we obtain
$$
(ad_{\lambda}^r M )\psi(k,x) = (-1)^r (ad_{D}^r q) \psi(k,x),
$$
where $ad_A B = A \circ B - B \circ A$ for any operators $A, B$.
Now, consider difference operator $\widetilde D$ given
by $\deg q$ iterations of the operation $ad$,
$$
\widetilde D = ad_D^{\deg q} q(k).
$$
As
$$
a(x)=(-1)^{\deg q} ad_{\lambda}^{\deg q} M
$$
becomes a polynomial in $x$, the function $\psi(k,x)$ is eigenfunction for
$\widetilde D$:
$$
\widetilde D \psi (k,x) = a(x) \psi (k,x),
$$
and therefore the commutativity relation holds:
$$
[D, \widetilde D]=0.
$$

It happens that difference operator $D$ allows simple construction
of Baker--Akhiezer function itself. This method was introduced by Chalykh in \cite{ChalBisp}
where such operators and BA functions for root systems and $A_{n,1}(m)$ deformation  were constructed. The formulas are as
follows

\begin{equation}
\label{BAformula}
\psi(k,x)= C(x) \left( D- \lambda(x) \right)^M \left( Q(k) e^{(k,x)}
\right)
\end{equation}
where the number of iterations $M=\sum_{\a\in A} m_\a$,
$Q(k)$ is the following polynomial in $k$
$$
Q(k)=
\prod_{\alpha\in A} \prod_{j=1}^{m_\alpha}
(k+j\alpha,\alpha) (k-j\alpha,\alpha),
$$
and $C(x)$ is a normalization function depending on $x$ variables
only. In the rational case such formulas for obtaining
Baker--Akhiezer functions  through applying differential
Calogero--Moser Hamiltonian  were found earlier by Berest
\cite{B}.

\subsection{Root systems}

Let $A=R=\{\alpha\}$ be a root system corresponding to semisimple Lie algebra
where we take exactly one of any pair of opposite roots.
Let function $m(\alpha)=m_\alpha$ be invariant with respect to the action
of the corresponding Weyl group.

\begin{predl}
For the system ${\cal R} = (R, m)$ there exists the
Baker--Akhiezer function.
\end{predl}

\begin{proof}
Essentially this statement contains in~\cite{VSCh}. More exactly, in ~\cite{VSCh}
it was shown the existence of function $\psi(k,x)$  having the desired form
(\ref{psi12}) but satisfying conditions
\begin{equation}\label{5}
\psi(k+s\alpha)=\psi(k-s\alpha),\quad \mbox{  }
\end{equation}
at $(k,\alpha)=0$ for all $\alpha\in R,
\ s=1,\dots,m_\alpha$.
It turns out that
$\psi(k,x)$  also satisfies~(\ref{22}). Indeed, we have to check that for
any $\alpha\in R$ and for any subsystem $R_+$ in $R\cup (-R)$ such that  $\alpha$ is edge
vector one has
\beq{6}
\prod_{\nad{\beta\in R_+}{\beta\ne \alpha}}\, \prod_{i=1}^{m_\beta}
(k+i\beta+s\alpha,\beta)
=
\prod_{\nad{\beta\in R_+}{\beta\ne \alpha}}\, \prod_{i=1}^{m_\beta}
(k+i\beta-s\alpha,\beta),
\eeq
for $(k,\a)=0$, $s=1,\dots,m_\alpha$.
We remark that condition that
 $\alpha$ is edge vector for $R_+$ means that $\alpha$ is a simple root with
 respect to $R_+$. We show that function
$\prod_{\nad{\beta\in R_+}{\beta\ne \alpha}} \prod_{i=1}^{m_\beta}
(k+i\beta,\beta)$ is symmetric with respect to $(\alpha,k)=0$,in
particular that the identity (\ref{6}) holds for arbitrary~$s$. Ä
Indeed, if $r_\alpha$ is reflection with respect to root $\alpha$
then
\begin{multline*}
r_\alpha
\prod_{\nad{\beta\in R_+}{\beta\ne \alpha}} \prod_{i=1}^{m_\beta}
(k+i\beta,\beta)
=
\prod_{\nad{\beta\in R_+}{\beta\ne \alpha}} \prod_{i=1}^{m_\beta}
(r_\alpha k+i\beta,\beta)
=\\=
\prod_{\nad{\beta\in R_+}{\beta\ne \alpha}} \prod_{i=1}^{m_\beta}
(k+i r_\alpha \beta, r_\alpha \beta)
=
\prod_{\nad{\gamma\in R_+}{\gamma\ne \alpha}} \prod_{i=1}^{m_\gamma}
(k+i\gamma,\gamma),
\end{multline*}
as for a simple root
 $\alpha$ the map
\hbox{$r_\alpha\!:\, R_+\setminus \alpha\to R_+\setminus \alpha$}
is a one-to-one correspondence not changing the multiplicity
function.
\end{proof}

In order to construct BA function let us first present the dual difference
operator $D$. For the root system $A_n$ this operator $D$ was found
by Ruijsenaars \cite{R1}, and for arbitrary root system operators $D$ were
introduced by Macdonald \cite{Macdonald}. For simplicity we will present here
formulas for all reduced root systems except $E_8, F_4, G_2$. The last
systems do not have so called minuscule coweight but we need
its existence for the formulas below. A minuscule coweight $\pi$ is such a coweight
that for any $\a\in R$ the scalar product $(\pi,\a)$ can take only
three values 0,1, and -1 at most.

For example, the root system $A_n$ consisting of the vectors $e_i-e_j$ in $\R^{n+1}$ has $n$ minuscule coweights
given by vectors $\pi_r= e_1+...+e_r$, where $1\le r \le n$.

So let us define following \cite{Macdonald} difference operator
$D_\pi$ by the formula
\begin{equation}
\label{D}
D_\pi = \sum_{\nad{\tau=w \pi}{w\in W}} \left( \prod_{\nad{\a\in
(R\cup (-R))}{(\a,\tau)=1}} \left(1-\frac{m_\a}{(\a,k)}\right) \right) T^\tau,
\end{equation}
where in the summation $W$ is the corresponding Weyl group, and
the operator $T^\tau$ is the operator which shifts a function
$f(k)$ to $f(k+\tau)$.
In the following way bispectral duality between Calogero--Moser--Sutherland and Ruijsenaars--Macdonald systems was established by Chalykh.

\begin{theorem} $(\cite{ChalBisp})$
Let ${\cal A} = (A, m)$ be a positive part of any reduced root
system of type $A,B,C,D$ or $E_6, E_7$ with invariant multiplicity
function. Let $\psi$ be the corresponding Baker--Akhiezer function
\mref{psi12}. Then the following two equations hold
$$
\left(\Delta - \sum_{\alpha\in A} \frac{m_\alpha (m_\alpha+1)
(\alpha,\alpha)}{\sinh^2 (\alpha,x)} \right)\psi = k^2 \psi,
$$
$$
D_\pi \psi = \sum_{w\in W} e^{(w \pi, x)} \psi,
$$
where  $D_\pi$ is the difference operator \mref{D} constructed for
the root system $\frac12 A^\vee$ with minuscule coweight $\pi$,
and $W$ is the corresponding Weyl group.
\end{theorem}
As it was shown in \cite{ChalBisp} the BA function can be
expressed by formula \mref{BAformula} where $D$ is operator given
by formula \mref{D} constructed from the dual system $\frac12
A^\vee$ which means that we consider the set of vectors
$\{\frac{\alpha}{(\a,\a)} \}$ instead of $\{\a\}$. And
$$
C(x)= \left(  \prod_{\a\in A} \left(\sum_{\nad{\tau=w \pi}{w\in W}} (\a,\tau) (\a,\a) e^{(\tau,x)}\right)^{m_\a}
\right)^{-1},
$$
where $\pi$ is a minuscule coweight for the root system
$\{\frac{\a}{(\a,\a)}\}$. As to $\lambda(x)$ it is given by the
formula
$$
\lambda(x) = \sum_{\nad{\tau=w \pi}{w\in W}} e^{(\tau,x)}.
$$

\subsection{Configuration $\pmb A_{n,1}(m)$}

The system $A_{n,1}(m)$ consists of the vectors $e_p-e_q$, $p<q,
\, p,q=1,\dots,n$, $m_{e_p-e_q}=m$, and vectors
$e_p-\sqrt{m}\,e_{n+1}$, $p=1,\dots,n$, $m_{e_p-\sqrt{m}\,
e_{n+1}}=1$. This configuration appeared in \cite{CFV1}. In
\cite{CFV2} it was shown that the corresponding rational and
trigonometric operators can be intertwined with Laplacian thus
they were algebraically integrable. In \cite{CFV3} it was shown
that rational version of the corresponding Schroedinger operator
admits the corresponding (symmetric) Baker--Akhiezer function. The
bispectral duality for the trigonometric version of this system as
well as existence of BA function in the sense of \cite{ChV1} was
obtained by Chalykh  in \cite{ChalBisp}.

\begin{predl}
There exists Baker--Akhiezer function for the system ${\cal A}=A_{n,1}(m)$.
\end{predl}

\begin{proof}
In the paper~\cite{ChalBisp} it was constructed a function
$\psi(k,x)$ of the form~(\ref{psi12}), satisfying
conditions~(\ref{5}) at $(k,\alpha)=0$ for all $ \alpha\in A, \
s=1,\dots,m_\alpha$. It happens that as in the case of root
systems 2.2.1, conditions~(\ref{5}) and~(\ref{22}) for the system
$A_{n,1}(m)$ are equivalent. Indeed, if $\alpha=e_p-e_q$, then
$\prod_{\beta\in A_+,\ \beta\ne \alpha} \prod_{i=1}^{m_\beta}
(k+i\beta,\beta)$ is symmetric with respect to $(\alpha,k)=0$.
Consider now $\alpha=e_p-\sqrt{m}\, e_{n+1}$. In order to
state~(\ref{5}) it is sufficient to check that  in any
two-dimensional plane $\pi$, $\pi\ni \alpha$ one has
\begin{equation}\label{7}
\prod_{\nad{\beta\in A_+\cap \pi}{\beta\ne \alpha}} \prod_{i=1}^{m_\beta}
(k+i\beta+\alpha,\beta)
=
\prod_{\nad{\beta\in A_+\cap \pi}{\beta\ne \alpha}} \prod_{i=1}^{m_\beta}
(k+i\beta-\alpha,\beta)
\end{equation}
at $k_p-\sqrt{m} k_{n+1}=0$.
There are two cases, either plane $\pi$ contains only one vector $\beta\in A_+$, $\beta\ne \alpha$, or $\pi$ contains two vectors $\beta_1$ and $\beta_2$.
In the first case  $(\alpha,\beta)=0$ and relation~(\ref{7}) holds. In the
second case condition $\beta\in A_+$ allows to set
$\beta_1=e_q-e_p$, $\beta_2=e_q-\sqrt{m}\, e_{n+1}$ or $\beta_1=e_p-e_q$, $\beta_2=\sqrt{m}\, e_{n+1}-e_q$ since $\alpha=e_p-\sqrt{m}\, e_{n+1}$ is
an edge vector. For the first choice identity
(\ref{7}) takes the form
\begin{multline*}
(k_q-k_p+1)\ldots(k_q-k_p+2m-1)(k_q-\sqrt{m}\, k_{n+1}+2m+1)
=\\=
(k_q-k_p+3)\ldots(k_q-k_p+2m+1)(k_q-\sqrt{m}\, k_{n+1}+1),
\end{multline*}
which is valid at $k_p=\sqrt{m}\, k_{n+1}$. The second choice also  gives
a valid identity.
\end{proof}

Let us present now the bispectral dual difference operator and the
formula for BA function both found by Chalykh in \cite{ChalBisp}.
The operator is given by the following formulae
$$
D=a_1 T_1 +\ldots + a_n T_n + a_{n+1} T_{n+1}^{\sqrt{m}},
$$
\begin{equation}
\label{Dchalykh}
a_i = \left( 1 - \frac{2}{k_i-\sqrt{m}k_{n+1}+ 1-m}\right)
\prod_{j\ne i}^n \left(1 - \frac{2m}{k_i-k_j}\right), \,\,
i=1,\ldots,n,
\end{equation}
$$
a_{n+1} = \frac{1}{m} \prod_{i=1}^n \left( 1 +  \frac{2m}{k_i-\sqrt{m}k_{n+1}+
1-m}\right),
$$
where the operators $T_i$ act on the functions $f(k)$ by
shifting the $i$th argument $k_i$ to $k_i+2$, and
$T^{\sqrt{m}}_{n+1} f(k_1,\ldots,k_{n+1})=
f(k_1,\ldots,k_{n+1}+2\sqrt{m}).
$

\begin{theorem} $(\cite{ChalBisp})$
Let $\psi(k,x)$ be the Baker--Akhiezer function for the system $A_{n,1}(m)$. Then $\psi(k,x)$ satisfies
the following difference equation
$$
D \psi(k,x)  = \lambda(x) \psi(k,x),
$$
where operator $D$ is given by formulas \mref{Dchalykh}, and
$$
\lambda(x)=e^{2x_1}+\ldots+e^{2x_n}+\frac{1}{m}e^{2\sqrt{m}x_{n+1}}.
$$
Also $\psi(k,x)$ itself can be expressed by formula
$$
\psi(k,x)= C(x) \left( D- \lambda(x) \right)^M \left( Q(k) e^{(k,x)}
\right)
$$
with
$$
C(x)=  \left(2^M M!  \prod_{i<j}^n  (e^{2 x_i} - e^{2 x_j})^m
\prod_{i=1}^n (e^{2 x_i}- e^{2 \sqrt{m}x_{n+1}}) \right)^{-1},
\quad M=m\frac{n(n-1)}{2}+n,
$$
$$
Q(k)= \prod_{i<j}^n\prod_{s=1}^m \left( (k_i-k_j)^2 - 4 s^2\right) \prod_{i=1}^n
\left( (k_i - \sqrt{m} k_{n+1})^2 - (m+1)^2 \right).
$$
\end{theorem}
\begin{remark}{When $m=1$ the system $A_{n,1}(m)$ coincides with the
root system $A_n$ with multiplicity $m=1$, and the operator $D$
degenerates to the corresponding Ruijsenaars--Macdonald operator
\mref{D} with coweight $\pi=e_1$.
}
\end{remark}

\section{Configuration \pmb{$C_n(l,m)$}}

This system consists of the following vectors in $\mathbb C^{n}$
depending on two parameters $l$, $m$: $\sqrt{2m+1}\, e_i$ with
multiplicity $m_i=l$, $i=1,\dots, n-1$; vector $\sqrt{2l+1}\,
e_{n}$ with multiplicity $m_n=m$; vectors $\frac{\sqrt{2m+1}}{2}\,
(e_i\pm e_j)$ with multiplicities $m_{ij}=\frac{2l+1}{2m+1}$,
$1\le i<j\le n-1$ (it is assumed that $\frac{2l+1}{2m+1}\in\mathbb
Z$); and vectors $\frac{\sqrt{2m+1}\,e_i\pm \sqrt{2l+1}\,
e_{n}}{2}$ with multiplicity $m_{in}=1$, $i=1,\dots,n-1$.

The configuration was introduced in \cite{CFV3} where the BA functions related
to rational potentials corresponding to  this system was under
investigations. For the trigonometric version related to $\cdva$ the intertwining operator to the pure Laplacian was constructed earlier in \cite{CFV2} (see also \cite{CFV1}).

We note at first that all the two-dimensional subsystems in $C_{n}
(l,m)$ have the form either of system $A_{2,1}(m)$ or of root
system or of subsystem $C_2(l,m)$. Ìû We have noticed already that
for a root system $R$ and for the system $A_{n,1}(m)$
identity~(\ref{6}) holds. It also holds for the system $C_2(l,m)$
and therefore for the system $C_{n}(l,m)$. Thus for system
$C_{n}(l,m)$, as well as for systems $R$, $A_{n,1}(m)$,
conditions~(\ref{22}) for Baker-Akhiezer function are equivalent
to simpler conditions~(\ref{5}).

Now we start constructing the BA function for the system
$C_{n}(l,m)$. The effective method we are going to use was found
by Chalykh~\cite{ChalBisp}. The method is based on finding
difference operator $D$ with special properties. Then BA function
$\psi(k,x)$ is obtained by multiple application of such operator
$D$ to some initial function $\vf_0$.

For the system $C_n(l,m)$ we define operator $D$ by the following formulas
\begin{equation}\label{8}
D=\sum_{i=1}^n a_i^+ T_i^+ + a_i^- T_i^-,
\end{equation}
where $T_i^\pm$ are difference operators which act as follows
\begin{gather*}
T_i^\pm f(k_1,\dots,k_i,\dots,k_n)=
f(k_1,\dots,k_i\pm\sqrt{2m+1},\dots,k_n),\\
i=1,\dots,n-1,
\\[2mm]
T_n^\pm f(k_1,\dots,k_n)=
f(k_1,\dots,k_n\pm \sqrt{2l+1})
\end{gather*}
The coefficients $a_i^\pm$ are functions of $k$ which are defined
by the formulas
$$
a_i^\pm =\prod_{j=1}^n a_{ij}^\pm,\qquad i=1,\dots,n,
$$
where
\begin{align*}
a_{ij}^\pm &=
\biggl(1-\frac{2l+1}{\pm \overline k_i+\overline k_j}\biggr)
\biggl(1-\frac{2l+1}{\pm \overline k_i-\overline k_j}\biggr),
\qquad 1\le i,j\le n-1,\  i\ne j,
\\[2mm]
a_{ii}^\pm &=
\frac1{2m+1}
\biggl(1-\frac{(2m+1)l}{\pm \overline k_i}\biggr),
\qquad i=1,\dots,n-1,
\\[2mm]
a_{in}^\pm &=
\biggl(1-\frac{2m+1}{\pm \overline k_i+\overline k_n-l+m}\biggr)
\biggl(1-\frac{2m+1}{\pm \overline k_i-\overline k_n-l+m}\biggr),
\\
&\hspace*{10cm}i=1,\dots,n-1,
\\[2mm]
a_{nj}^\pm &=
\biggl(1-\frac{2l+1}{\pm \overline k_n+\overline k_j+l-m}\biggr)
\biggl(1-\frac{2l+1}{\pm \overline k_n-\overline k_j+l-m}\biggr),
\\
&\hspace*{10cm}j=1,\dots,n-1,
\\[2mm]
a_{nn}^\pm &=
\frac1{2l+1}
\biggl(1-\frac{(2l+1)m}{\pm \overline k_n}\biggr).
\end{align*}
In the above formulas and throughout this section we use notation $\overline k_i=\sqrt{2m+1}\,
k_i$ for  $i=1,\dots,n-1$, and $\overline k_n=\sqrt{2l+1}\, k_n$.

\begin{remark}{When $m=l$ the system $\cnm$ becomes the
root system $C_n$ consisting of the vectors $\sqrt{2m+1}e_i$ with multiplicities $m$ and vectors $\frac{\sqrt{2m+1}}2(e_i \pm e_j)$ with multiplicity $1$. Then operator \mref{8} is a  $\frac1{2m+1}$ multiple of the corresponding  Macdonald operator \mref{D} written for the root system $B_n=\frac12 C_n^\vee$ consisting
of the vectors $\frac1{\sqrt{2m+1}} e_i$ with multiplicity $m$, $\frac1{\sqrt{2m+1}}(e_i \pm e_j)$ with multiplicity $1$, and the minuscule coweight $\pi=\sqrt{2m+1} e_1$.
}
\end{remark}

The next step is to prove invariance of the space $V$ of
holomorphic functions $f(k)$ satisfying
\beq{55}
f(k+s\a) = f(k - s\a)  \quad \mbox{ at } (k,\a)=0
\eeq
for $s=1,\ldots, m_\a$, for all $\a\in \cnm$,
 under the action
of operator (\ref{8}). Notice that for the system $C_n(l,m)$
conditions~(\ref{55}) can be rewritten in the following form. For
$\alpha=\sqrt{2m+1}\, e_i, i\le n-1$, and $\a=\sqrt{2l+1}\, e_n$
\begin{equation}\label{9}
(T_i^+)^s f=(T_i^-)^s f
\quad \mbox{at } \overline k_i=0,
\quad i=1,\dots,n-1, \ s\le l; \mbox{  and } \  i=n, s\le m.
\end{equation}
For $\alpha=\frac{\sqrt{2m+1}}{2}\, (e_i-e_j)$
\begin{equation}\label{10}
\begin{gathered}
(T_i^+)^s f=(T_j^+)^s f
\\
\mbox{at }\  \overline k_i-\overline k_j=0,
\quad i,j=1,\dots,n-1,\ s=1,\ldots,\frac{2l+1}{2m+1},
\end{gathered}
\end{equation}
or equivalently
\begin{equation}\label{10'}
\begin{gathered}
(T_i^-)^s f=(T_j^-)^s f
\\
\mbox{at }\  \overline k_i-\overline k_j=0,
\quad i,j=1,\dots,n-1, \ s=1,\ldots,\frac{2l+1}{2m+1}.
\end{gathered}
\tag{$\ref{10}'$}
\end{equation}
For $\alpha=\frac{\sqrt{2m+1}}{2}\, (e_i+e_j)$
\begin{equation}\label{11}
\begin{gathered}
(T_i^+)^s f=(T_j^-)^s f
\\ \mbox{at }\  \overline k_i+\overline k_j=0,
\quad i,j=1,\dots,n-1,\ s=1,\ldots,\frac{2l+1}{2m+1},
\end{gathered}
\end{equation}
or equivalently
\begin{equation}\label{11'}
\begin{gathered}
(T_i^-)^s f=(T_j^+)^s f
\\
\mbox{at }\  \overline k_i+\overline k_j=0,
\quad i,j=1,\dots,n-1,\ s=1,\ldots,\frac{2l+1}{2m+1}.
\end{gathered}
\tag{$\ref{11}'$}
\end{equation}
For $\alpha=\frac{\sqrt{2m+1}\, e_i-\sqrt{2l+1}\, e_n}{2}$
\begin{equation}\label{12}
T_i^+ f=T_n^+ f
\qquad \mbox{at } \overline k_i-\overline k_n-l+m=0,
\quad i=1,\dots,n-1,
\end{equation}
or equivalently
\begin{equation}\label{12'}
T_i^- f=T_n^- f
\qquad \mbox{at } \overline k_i-\overline k_n+l-m=0,
\quad i=1,\dots,n-1.
\tag{$\ref{12}'$}
\end{equation}
Finally, for the case $\alpha=\frac{\sqrt{2m+1}\,
e_i+\sqrt{2l+1}\, e_n}{2}$ conditions~(\ref{55}) may be represented
as
\begin{equation}\label{13}
T_i^+ f=T_n^- f
\qquad \mbox{at } \overline k_i+\overline k_n-l+m=0,
\quad i=1,\dots,n-1,
\end{equation}
and also
\begin{equation}\label{13'}
T_i^- f=T_n^+ f
\qquad \mbox{at } \overline k_i+\overline k_n+l-m=0,
\quad i=1,\dots,n-1.
\tag{$\ref{13}'$}
\end{equation}

The validity of the transformation from the form~(\ref{55}) to the
form (\ref{9})--(\ref{13}) can be simply established. For example,
consider condition~(\ref{55}) for $\alpha=\frac{\sqrt{2m+1}\,
e_i+\sqrt{2l+1}\, e_n}{2}$. Obviously it can be written as
$$
(T_i^+ - T_n^-)
f\biggl(k+\frac{-\sqrt{2m+1}\, e_i+\sqrt{2l+1}\, e_n}{2}\biggr)=0,
\qquad \mbox{at }\
\overline k_i+\overline k_n=0.
$$
We are left to point out that the set
$$
\biggl\{
k+\frac{-\sqrt{2m+1}\, e_i+\sqrt{2l+1}\, e_n}{2}
\ \Bigr| \quad
\overline k_i+\overline k_n=0
\biggr\}
$$
is given by the equation $\overline k_i+\overline k_n+m-l=0$. Thus
we arrive to the record~(\ref{13}). Representing
condition~(\ref{55}) in the form
$$
(T_i^- - T_n^+)
f\biggl(k+\frac{\sqrt{2m+1}\, e_i-\sqrt{2l+1}\, e_n}{2}\biggr)=0,
\qquad \mbox{at }
\overline k_i+\overline k_n=0.
$$
we get record~(\ref{13'}). The form~(\ref{12}) is obtained
analogously. The equivalence of the conditions
(\ref{9})--(\ref{11}) to the corresponding conditions (\ref{55}) is
obvious.

\begin{predl}\label{predl4} Let $D$ be operator $(\ref{8})$, let $f(k_1,\dots,k_n)$
be any holomorphic function  satisfying conditions
$(\ref{9})$--$(\ref{13})$. Then function $Df(k_1,\dots,k_n)$ is
also holomorphic.
\end{predl}

\begin{proof}
In principle function $Df(k_1,\dots,k_n)$ could have singularities at the hyperplanes
where operator $D$ is singular. We will show that this doesn't happen by
subsequent  consideration of singularities of operator $D$.

a) $k_i=0$, $i=1,\dots,n$. We collect terms in $Df(k_1,\dots,k_n)$
which are singular at $k_i=0$.
$$
Df=\sum_{j=1}^n a_j^+ T_j^+ (f)
+ a_j^- T_j^- (f)
=
-\frac{\epsilon}{\overline k_i}
\biggl(\prod_{j\ne i} a_{ij}^+ T_i^+ f-
\prod_{j\ne i} a_{ij}^- T_i^- f\biggr)
+f_i(k),
$$
where $\epsilon=l$ for $i<n$ and $\epsilon=m$ for $i=n$; the
functions $f_i(k)$ are holomorphic at $\overline k_i=0$. We note
that $a_{ij}^+=a_{ij}^-$ at $\overline k_i=0$, therefore
$a_{ij}^+=a_{ij}^- + \overline k_i h_{ij}(k)$ where $h_{ij}(k)$
are holomorphic at $\overline k_i=0$, and we obtain relation
$$
\sum_{j=1}^n a_j^+ T_j^+ f + a_j^- T_j^- f=
-\Bigl(\epsilon\prod_{j\ne i} a_{ij}^+ \Bigr) \frac{1}{\overline k_i}
(T_i^+ f - T_i^- f) +\widetilde f_i(k),
$$
where $\widetilde f_i(k)$ is holomorphic at $\overline k_i=0$. Thus because
of conditions~(\ref{9}) the function $D f$ is non-singular at
$\overline k_i=0$.

b)
$\overline k_i-\overline k_j=0$, $i,j=1,\dots,n-1$.
For appropriate functions $f_{ij}$, $\widetilde f_{ij}$,
$\dtilde f_{ij}$  holomorphic at
$\overline k_i=\overline k_j$ the following chain of equalities takes place
{\normalsize
\begin{multline*}
Df=a_i^+\, T_i^+ f + a_j^+\, T_j^+ f  +
a_i^-\, T_i^- f + a_j^-\, T_j^- f + f_{ij}
=\\[2mm] =
-\frac{2l+1}{\overline k_i - \overline k_j}
\biggl(1-\frac{2l+1}{\overline k_i + \overline k_j}\biggr)
\,\frac{1}{2m+1}\,
\biggl(1-\frac{(2m+1)l}{\overline k_i}\biggr)
\prod_{s\ne i,j} a_{is}^+\, T_i^+ f
-\\[2mm] -
\frac{2l+1}{\overline k_j - \overline k_i}
\biggl(1-\frac{2l+1}{\overline k_i + \overline k_j}\biggr)
\,\frac{1}{2m+1}\,
\biggl(1-\frac{(2m+1)l}{\overline k_j}\biggr)
\prod_{s\ne i,j} a_{js}^+\, T_j^+ f
+\\[2mm] +
\frac{2l+1}{\overline k_i - \overline k_j}
\biggl(1+\frac{2l+1}{\overline k_i + \overline k_j}\biggr)
\,\frac{1}{2m+1}\,
\biggl(1+\frac{(2m+1)l}{\overline k_i}\biggr)
\prod_{s\ne i,j} a_{is}^-\, T_i^- f
+\\[2mm] +
\frac{2l+1}{\overline k_j - \overline k_i}
\biggl(1+\frac{2l+1}{\overline k_i + \overline k_j}\biggr)
\,\frac{1}{2m+1}\,
\biggl(1+\frac{(2m+1)l}{\overline k_j}\biggr)
\prod_{s\ne i,j} a_{js}^-\, T_j^- f
+\widetilde f_{ij}
=\\[2mm] =
-(2l+1)
\biggl(1-\frac{2l+1}{\overline k_i + \overline k_j}\biggr)
\,\frac{1}{2m+1}\,
\biggl(1-\frac{(2m+1)l}{\overline k_i}\biggr)
\prod_{s\ne i,j} a_{is}^+
\, \cdot \,
\frac{1}{\overline k_i - \overline k_j}
(T_i^+ f - T_j^+ f)
+\\+
(2l+1)
\biggl(1+\frac{2l+1}{\overline k_i + \overline k_j}\biggr)
\,\frac{1}{2m+1}\,
\biggl(1+\frac{(2m+1)l}{\overline k_i}\biggr)
\prod_{s\ne i,j} a_{is}^+
\, \cdot \,
\frac{1}{\overline k_i - \overline k_j}
(T_i^- f - T_j^- f)
+\dtilde f_{ij},
\end{multline*}

} \noindent as one has $a_{is}^\pm=a_{js}^\pm$ at $k_i=k_j$ for
$s\ne i,j$. Thus because of conditions~(\ref{10}), (\ref{10'})
function $D f$ has no singularities at $\overline k_i - \overline
k_j=0$. Further it is easy to see the invariance of operator $D$
under reflections around $\overline k_j=0$, $j=1,\dots,n$.  But
the hyperplane $\overline k_i - \overline k_j=0$ is mapped to
$\overline k_i + \overline k_j=0$ under such a reflection.
Therefore $Df$ is non-singular also at the hyperplanes $\overline
k_i + \overline k_j=0$, $i,j=1,\dots,n-1$.

\vspace{2mm}

We are left to analyze possible singularities of function $Df$ at
the hyperplanes $\overline k_i \pm \overline k_n \pm (l-m)=0$,
$i=1,\dots,n-1$. Because of mentioned symmetry of the operator $D$
it is enough to restrict considerations to the hyperplanes
$\overline k_i - \overline k_n + l-m=0$.

c) $\overline k_i - \overline k_n + l-m=0$, $i=1,\dots,n-1$.
The coefficients of operator $D$ which are singular at this hyperplane are
$a_i^-$ è $a_n^-$. Ìû We have
\begin{multline*}
Df=a_i^-\, T_i^- f +  a_n^-\, T_n^- f + f_{in}
=\\=
\frac1{2m+1}
\biggl( 1+\frac{(2m+1)l}{\overline k_i}\biggr)
\biggl( 1+\frac{2m+1}{\overline k_i+\overline k_n +l-m}\biggr)
\times\hfill\null
\\  \null\hfill\times
\prod_{j\ne i,n} a_{ij}^-
\biggl(-\frac{2m+1}{-\overline k_i+\overline k_n -l+m}\biggr)
T_i^- f+
\\[2mm] +
\frac1{2l+1}
\biggl( 1+\frac{(2l+1)m}{\overline k_n}\biggr)
\biggl( 1+\frac{2l+1}{\overline k_n+\overline k_i -l+m}\biggr)
\times\hfill\null
\\\times
\prod_{j\ne i,n} a_{nj}^-
\biggl(-\frac{2l+1}{-\overline k_n+\overline k_i +l-m}\biggr)
T_n^- f
+\widetilde f_{in},
\end{multline*}
where $f_{in}$, $\widetilde f_{in}$ are some functions which are holomorphic
at $\overline k_i-\overline k_n +l-m=0$. Obviously one has
$ a_{ij}^-= a_{nj}^-$, $j\ne i,n$ at $\overline k_i-\overline k_n +l-m=0$.
Moreover, one has
\begin{multline*}
\biggl( 1+\frac{(2m+1)l}{\overline k_i}\biggr)
\biggl( 1+\frac{2m+1}{\overline k_i+\overline k_n +l-m}\biggr)
=\\[2mm] =
\frac{\overline k_i+(2m+1)l}{\overline k_i}
\cdot
\frac{\overline k_i+\overline k_n+l+m+1}{\overline k_i+\overline k_n+l-m}
=
\\[2mm]
=
\frac{\overline k_n+(2l+1)m}{\frac12 (\overline k_i+\overline k_n-l+m)}
\cdot
\frac{\overline k_i+\overline k_n+l+m+1}{2\overline k_n}=
\\[2mm] =
\biggl( 1+\frac{(2l+1)m}{\overline k_n}\biggr)
\biggl( 1+\frac{2l+1}{\overline k_n+\overline k_i -l+m}\biggr).
\end{multline*}
at this hyperplane. Therefore we can extend equality for $D f$ as follows
\begin{multline*}
D f=
\biggl( 1+\frac{(2m+1)l}{\overline k_i}\biggr)
\biggl( 1+\frac{2m+1}{\overline k_i+\overline k_n +l-m}\biggr)
\times\\\times
\prod_{j\ne i,n}  a_{ij}^-
\,\cdot\,
\frac{1}{\overline k_i-\overline k_n +l-m}
(T_i^- f - T_n^- f)
+ \dtilde f_{in}
\end{multline*}
for some function $\dtilde f_{in}$ holomorphic at
$\overline k_i-\overline k_n +l-m=0$. Because of~(\ref{12'}) the function
$D f$ is non-singular at $\overline k_i-\overline k_n +l-m=0$. Thus the proposition
is fully proved.
\end{proof}

\begin{predl}\label{predl5}
For any holomorphic function $f(k_1,\dots,k_n)$ satisfying conditions $(\ref{9})$--$(\ref{13})$
the function
$Df(k_1,\dots,k_n)$ also satisfies conditions  $(\ref{9})$--$(\ref{13})$
 if $D$ is operator $(\ref{8})$.
\end{predl}

\begin{proof} We consider different hyperplanes and subsequently show that
operator $D$ keeps axiomatics at any of the hyperplanes.\\

a) $\pi_i = \{k_i=0\}$, $1\le i \le n$.\\
We have
\begin{multline}\label{ki}
(\tips-\tims)Df = (\tips-\tims)\sum_{j=1}^n(\ajp\tjp+\ajm\tjm)f=
\\=
\sum_{j\ne i} \left(\tips(\ajp)\tjp\tips f - \tims(\ajp)\tjp\tims f\right) +
\\+
\sum_{j\ne i} \left(\tips(\ajm)\tjm\tips f - \tims(\ajm)\tjm\tims f\right) +
\\+
\tips(\aip)\tipspo f - \tims(\aim)\timspo f +
\tips(\aim)\tipsmo f - \tims(\aip)\timsmo f
\end{multline}
If $j\ne i$ then functions $\ajpm$ are invariant with respect to
reflection $s_i$ around hyperplane $\pi_i$. Therefore
$\tips(\ajpm)|_{\pi_i} = \tims(\ajpm)|_{\pi_i}$. Because
$s_i(\aip)=\aim$ we get $s_i(\tips(\aip)) = \tims(\aim)$ and in
particular $\tips(\aip)|_{\pi_i}=\tims(\aim)|_{\pi_i}$. Thus the
right-hand side of (\ref{ki}) can be rewritten in the form
\begin{multline*}
\sum_{j\ne i} \tips(\ajp)\tjp\left(\tips  - \tims \right)f +
\sum_{j\ne i} \tips(\ajm)\tjm\left(\tips  - \tims \right)f +
\\+
\tips(\aip)\big(\tipspo  - \timspo\big) f +
\tips(\aim)\big(\tipsmo  - \timsmo\big) f.
\end{multline*}
Because o conditions (\ref{9}) at $s<m_i$ everything is proven. For
$s=m_i$ we are left to notice that $\tips(\aipm)|_{\pi_i}=0$.\\

b) $\pi_{ij} = \{k_i=k_j\}$, $1\le i<j < n$.\\
We have
\begin{multline}\label{kij}
(\tips-\tjps)Df = (\tips-\tjps)\sum_{q=1}^n(\aqp\tqp+\aqm\tqm)f=
\\[1mm]=
\sum_{q\ne i,j} \big(\tips(\aqp)\tqp\tips f - \tjps(\aqp)\tqp\tjps f\big) +
\\[1mm]+
\sum_{q\ne i,j} \big(\tips(\aqm)\tqm\tips f - \tjps(\aqm)\tqm\tjps f\big) +
\\[1mm]+
\big(\tips(\aip)\tipspo f - \tjps(\ajp)\tjpspo f\big) +
\\[1mm]+
\big(\tips(\aim)\tipsmo f - \tjps(\ajm)\tjpsmo f\big)+
\\[1mm]+
\big(\tips(\ajp)\tips\tjp f - \tjps(\aip)\tjps\tip f\big) +
\\[1mm]+
\big(\tips(\ajm)\tips\tjm f - \tjps(\aim)\tjps\tim f\big).
\end{multline}
We show that sum (\ref{kij}) vanishes at the hyperplane
$\pi_{ij}$. For $q\ne i,j$ the functions $\aqpm$ are invariant with respect to reflection $s_{ij}$ around hyperplane $k_i=k_j$. ÑTherefore
$\tips(\aqpm)|_{\pi_{ij}} =
\tjps(\aqpm)|_{\pi_{ij}}$. Ò
Because $s_{ij}(\aipm)=\ajpm$ we get
$$
s_{ij}(\tips(\aipm)) = \tjps(\ajpm), \qquad
s_{ij}(\tjps(\aipm)) =
\tips(\ajpm),
$$
and in particular
$$
\tips(\aipm)|_{\pi_{ij}}=\tjps(\ajpm)|_{\pi_{ij}}, \qquad
\tjps(\aipm)|_{\pi_{ij}}=\tips(\ajpm)|_{\pi_{ij}}.
$$
Totally we conclude that the right-hand side of (\ref{kij}) can be rewritten
as
\begin{multline}\label{kij2}
\sum_{q\ne i,j} \tips(\aqp)\tqp\left(\tips  - \tjps \right)f +
\sum_{q\ne i,j} \tips(\aqm)\tqm\left(\tips  - \tjps \right)f +
\\+
\tips(\aip)\left(\tipspo  - \tjpspo\right) f +
\tips(\aim)\left(\tipsmo  - \tjpsmo\right) f +
\\
+
\tips(\ajp)\tip\tjp\left(\tipsmo  - \tjpsmo\right) f +
\tips(\ajm)\tim\tjm\left(\tipspo  - \tjpspo\right) f.
\end{multline}
Because of conditions (\ref{10}) the first two sums in
(\ref{kij2}) equal zero. As the shifts along the vectors $\overline
e_i+\overline e_j$, $- \overline e_i - \overline e_j$ do not change $\overline k_i -
\overline k_j$, the left four terms at  $s<m_{ij}$ also vanish because
of (\ref{10}). If $s=m_{ij}$ then this is correct if we recall
that
${\tip}^{m_{ij}}(\aip) = {\tip}^{m_{ij}}(\ajm) = 0 $ at $k\in \pi_{ij}$.\\

c) $\pi_{in} = \{\overline k_n - \overline k_i +l - m =0 \}$, $1\le i < n$.\\
We have
\begin{multline}\label{kni}
(\tnp-\tip)Df = (\tnp-\tip)\sum_{q=1}^n(\aqp\tqp+\aqm\tqm)f=
\\=
\sum_{q\ne i,n} \left(\tnp(\aqp)\tqp\tnp f - \tip(\aqp)\tqp\tip f\right) +
\\+
\sum_{q\ne i,n} \left(\tnp(\aqm)\tqm\tnp f - \tip(\aqm)\tqm\tip f\right) +
\\+
\left(\tnp-\tip\right)\left(\anp\tnp +
\anm\tnm+\aip\tip+\aim\tim\right)f.
\end{multline}
We notice that both sums in (\ref{kni}) vanish like in the case b) because
$\tnp(\aqpm)=\tip(\aqpm)$. Indeed,
\begin{multline*}
\tnp(\aqpm)=\prod_{t\ne n} a_{qt}^\pm \tnp(a_{qn}^\pm)
=
\\=
 \prod_{t\ne n,i} a_{qt}^\pm a_{qi}^\pm\tnp\left(1-\frac{2m+1}{\pm\overline k_q+\overline
k_n-l+m}\right)\left(1-\frac{2m+1}{\pm\overline k_q-\overline
k_n-l+m}\right) =
\\=
\prod_{t\ne n,i} a_{qt}^\pm\left(1-\frac{2l+1}{\pm\overline k_q+\overline
k_i}\right) \left(1-\frac{2l+1}{\pm\overline k_q-\overline k_i}\right)\times
\\\times
\left(1-\frac{2m+1}{\pm\overline k_q+\overline
k_n+l+m+1}\right)\left(1-\frac{2m+1}{\pm\overline k_q-\overline
k_n-3l+m-1}\right).
\end{multline*}
Analogously,
\begin{multline*}
\tip(\aqpm) =
\prod_{t\ne n,i} a_{qt}^\pm\left(1-\frac{2m+1}{\pm\overline k_q+\overline
k_n-l+m}\right) \left(1-\frac{2m+1}{\pm\overline k_q-\overline k_n-l+m}\right)
\times
\\
\times
\left(1-\frac{2l+1}{\pm\overline k_q+\overline
k_i+2m+1}\right)\left(1-\frac{2l+1}{\pm\overline k_q-\overline
k_i-2m-1}\right).
\end{multline*}
It is easy to check that if $k\in \pi_{in}$ then one has
\begin{multline*}
\left(1-\frac{2l+1}{\pm\overline k_q+\overline
k_i}\right) \left(1-\frac{2l+1}{\pm\overline k_q-\overline k_i}\right)
\left(1-\frac{2m+1}{\pm\overline k_q+\overline
k_n+l+m+1}\right)\times
\\
\null\hfill\times
\left(1-\frac{2m+1}{\pm\overline k_q-\overline
k_n-3l+m-1}\right)
=
\\[2mm]
=
\left(1-\frac{2m+1}{\pm\overline k_q+\overline
k_n-l+m}\right) \left(1-\frac{2m+1}{\pm\overline k_q-\overline k_n-l+m}\right)
\times
\hfill\null
\\\times
\left(1-\frac{2l+1}{\pm\overline k_q+\overline
k_i+2m+1}\right)
\left(1-\frac{2l+1}{\pm\overline k_q-\overline
k_i-2m-1}\right).
\end{multline*}
Thus the right-hand side of (\ref{kni}) is simplified to the following expression
\begin{multline}\label{kni2}
\tnp(\anp){\tnp}^2 f -\tip(\aip){\tip}^2 f +
\tnp(\aim)\tnp\tim f -\tip(\anm)\tip\tnm f +
\\+
\big(\tnp(\aip)-\tip(\anp)\big)\tip\tnp f + \big(\tnp(\anm)-\tip(\aim)\big)f.
\end{multline}
We note that
$$
\tnp(a_{ni}^+)=\tip(a_{in}^+)=\tnp(a_{in}^i)=\tip(a_{ni}^-)=0
$$
at $k\in\pi_{in}$. We are left to check that
\beq{ost1}
\tnp(\aip)=\tip(\anp),
\eeq
\beq{ost2}
\tnp(\anm)=\tip(\aim).
\eeq
We note that if $t\ne i,n$ then
$$
\tnp(a_{it}^+) = a_{it}^+ = a_{nt}^+ = T_i^+(a_{nt}^+).
$$
Therefore condition (\ref{ost1}) is reduced to the condition $a_{ii}^+\tnp(a_{in}^+) =
a_{nn}^+\tip(a_{ni}^+)$, or
\begin{multline*}
\frac1{2m+1}\left(1-\frac{(2m+1)l}{\overline k_i}\right)
\left(1-\frac{2m+1}{\overline k_i+\overline
k_n+l+m+1}\right)
\times\\
\null\hfill\times
\left(1-\frac{2m+1}{\overline k_i-\overline
k_n-3l+m-1}\right) =
\\[2mm]=
\frac1{2l+1} \left(1-\frac{(2l+1)m}{\overline k_n}\right)
\left(1-\frac{2l+1}{\overline k_n+\overline
k_i+l+m+1}\right)
\times \hfill\null
\\
\null\hfill\times
\left(1-\frac{2l+1}{\overline k_n-\overline
k_i+l-3m-1}\right),
\end{multline*}
which is valid. We are left to check condition (\ref{ost2}). We note that
if  $t\ne i,n$ then
\begin{multline*}
\tnp(a_{nt}^-) = \tnp(1-\frac{2l+1}{-\overline k_n+\overline
k_t+l-m})(1-\frac{2l+1}{-\overline k_n-\overline k_t+l-m})=
\\=
(1-\frac{2l+1}{-\overline k_n+\overline
k_t-l-m-1})(1-\frac{2l+1}{-\overline k_n-\overline k_t-l-m-1})=
\\=
(1-\frac{2l+1}{-\overline k_i+\overline
k_t-2m-1})(1-\frac{2l+1}{-\overline k_i-\overline k_t-2m-1})=
\\=
\tip(1-\frac{2l+1}{-\overline k_i+\overline
k_t})(1-\frac{2l+1}{-\overline k_i-\overline k_t})= \tip(a_{it}^-).
\end{multline*}
Therefore identity (\ref{ost2}) is reduced to the following
$$
\tnp(a_{nn}^- a_{ni}^-)=\tip(a_{ii}^- a_{in}^-).
$$
Substituting the corresponding expressions we get
\begin{multline*}
\frac1{2l+1}\left(1+\frac{(2l+1)m}{\overline k_n +2l+1}\right)
\left(1-\frac{2l+1}{-\overline k_n+\overline
k_i-l-m-1}\right)\times
\\[2mm]\times
\left(1-\frac{2l+1}{-\overline k_n-\overline
k_i-l-m-1}\right) =
\frac1{2m+1} \left(1+\frac{(2m+1)l}{\overline k_i+2m+1}\right)\times
\\[2mm]\times
\left(1-\frac{2m+1}{-\overline k_i+\overline
k_n-l-m-1}\right)
\left(1-\frac{2m+1}{-\overline k_i-\overline
k_n-l-m-1}\right),
\end{multline*}
equivalently,
\begin{multline*}
\left(1+\frac{(2l+1)m}{\overline k_n+2l+1}\right)\left(1+\frac{2l+1}{\overline k_n+\overline
k_i+l+m+1}\right) =
\\[2mm]=
\left(1+\frac{(2m+1)l}{\overline k_i+2m+1}\right)\left(1+\frac{2m+1}{\overline k_i+\overline
k_n+l+m+1}\right),
\end{multline*}
which is valid for $k\in\pi_{in}$.\\

d) The fact that axiomatics at the hyperplanes $k_i+k_j=0$, $\overline k_n
+ \overline k_i +l-m=0$, \ $i,j = 1,\ldots,n-1$ is preserved can be checked analogously
to the cases b) and c) correspondingly.\\
Thus the proposition is proved.
\end{proof}

Now we are ready to construct Baker--Akhiezer function
$\psi(k,x)$. We define sequence of functions $\varphi_i(k,x)$ by
the following formulas. Let
\begin{equation}
\label{aa}
\varphi_0=\prod_{\alpha\in \cnm} \prod_{s=1}^{m_\alpha}
(k+s\alpha,\alpha) (k-s\alpha,\alpha) e^{(k,x)}
\end{equation}
More explicitly we have
\begin{multline*}
\varphi_0 = a
\prod_{i=1}^{n-1}\prod_{s=1}^l (\overline k_i^2-s^2 (2m+1)^2)
\prod_{s=1}^m (\overline k_n^2-s^2(2l+1)^2)\times\\
\prod_{i=1}^{n-1}\left( (\overline k_i + \overline k_n)^2 - (m+l+1)^2 \right)
\left( (\overline k_i - \overline k_n)^2 - (m+l+1)^2 \right)
\times\\
\prod_{i<j}^{n-1} \prod_{s=1}^{\frac{2l+1}{2m+1}} \left( (\overline k_i+\overline k_j)^2
- s^2 (2m+1)^2 \right)\left( (\overline k_i-\overline k_j)^2 - s^2 (2m+1)^2\right) e^{(k,x)}
\end{multline*}
where
$$
a=   2^{2(1-n)(2+(n-2)\frac{2l+1}{2m+1})}.
$$
Then we define
\beq{bb}
\varphi_{i+1}=\biggl(
D-\frac{2}{2m+1}\sum_{j=1}^{n-1} \cosh \sqrt{2m+1} \, x_j -
\frac{2}{2l+1} \cosh \sqrt{2l+1} \, x_n
\biggr) \varphi_i.
\eeq
It turns out that at the step
\beq{summ}
M = \sum_{\a\in \cnm} m_\a = (2+l)(n-1)+m+(n-1)(n-2)\frac{2l+1}{2m+1}
\eeq
one gets the BA function. Before formulating the theorem let us introduce
abbreviations  $\overline x_i = \sqrt{2m+1} x_i$
 for $i=1,\ldots, n-1$, and $\overline x_n = \sqrt{2l+1}x_n$.

\begin{theorem}\label{teorCsyst}
ÔThe Baker--Akhiezer function is given by formula
$$
\psi(k,x)=c^{-1}(x)\varphi_{M},
$$
where $\varphi_{M}$ is defined by formulas $(\ref{aa}),
(\ref{bb}), \mref{summ}$, and
\begin{multline*}
c(x)=
M !
\ (e^{\overline x_n}-e^{-\overline x_n})^m
\times
\\
\times
\prod_{i=1}^{n-1} (e^{\overline x_i}-e^{-\overline x_i})^l
\prod_{i<j}^{n-1}    (e^{\overline x_i-\overline x_j}-e^{\overline x_j-\overline x_i})^\frac{2l+1}{2m+1}
(e^{\overline x_i+\overline x_j}-e^{-\overline x_i-\overline x_j})^\frac{2l+1}{2m+1}
\times
\\
\times
\prod_{i=1}^{n-1}    (e^{\overline x_i-\overline x_n}-e^{\overline x_n-\overline x_i})
(e^{\overline x_i+\overline x_n}-e^{-\overline x_i-\overline x_n}).
\end{multline*}
\end{theorem}

\begin{proof}
For the function $\varphi_0$ the axiomatic conditions
(\ref{2'}) are clearly satisfied. Therefore in view of propositions \ref{predl4}, \ref{predl5} these conditions would also hold for all
$\varphi_i (k,x)$, and
$\varphi_i (k,x)=P_i(k,x)\, e^{(k,x)}$ where $P_i$ is a polynomial in $k$.
Further we use induction to find the highest term $P_i^0 (k,x)$ of the polynomial
$P_i$.

By definition for any $s \in \N$ we have
\begin{multline}
\label{otn}
(P_{s+1}^0+\mbox{ lower order terms in } P_{s+1}) e^{(k,x)}=
\\
=\biggl(
D - \frac{1}{2m+1} \sum_{j=1}^{n-1} e^{\overline x_j} -
\frac{1}{2m+1} \sum_{j=1}^{n-1} e^{-\overline x_j} -
\frac{1}{2l+1}  e^{\overline x_n} -
\frac{1}{2l+1}  e^{-\overline x_n}
\biggr)\times
\\
\times
(P_{s}^0+\mbox{lower order terms in } P_s) e^{(k,x)}.
\end{multline}
In order to get formulas for $P_{s+1}^0$ we represent the
right-hand side of (\ref{otn}) as a fraction of two polynomials.
In the denominator of (\ref{otn}) it will be polynomial
\begin{multline*}
N=\prod_{i<j}^{n-1} (\overline k_i+ \overline k_j) (\overline k_i- \overline k_j)
\times
\\
\times
\prod_{i=1}^{n-1} (\overline k_i+ \overline k_n-l+m)
(\overline k_i- \overline k_n-l+m)
(-\overline k_i+ \overline k_n-l+m)
(-\overline k_i- \overline k_n-l+m)
\prod_{i=1}^{n} \overline k_i.
\end{multline*}
We continue equality (\ref{otn}) using formulas for the coefficients of operator $D$ given by (\ref{8}). We introduce notation $[Q(k)]^0$ for the highest
homogeneous part of polynomial $Q(k)$. Let $N^1$ denote homogeneous component
of polynomial $N$ of degree $\deg N - 1$. Then up to the lower terms we have
$$
\bigg(D-
\frac{1}{2m+1} \sum_{j=1}^{n-1} (e^{\overline x_j}+e^{-\overline x_j})-
\frac{1}{2l+1}  (e^{\overline x_n}+e^{-\overline x_n})
\bigg)
(P_s^0+P_s^1+\ldots) e^{(k,x)}
=
$$
\begin{multline*}
= \frac1N \Biggl\{
\frac{1}{2m+1} \sum_{i=1}^{n-1}
\bigg(
N^0+N^1 -
\Big[ N\Big(\sum_{j\ne i}   \left(
\frac{2l+1}{\overline k_i+\overline k_j} + \frac{2l+1}{\overline k_i-\overline
k_j}\right)
+\frac{(2m+1)l}{\overline k_i} +
\\
+\frac{2m+1}{\overline k_i+\overline k_n-l+m}+
\frac{2m+1}{\overline k_i-\overline k_n-l+m}   \Big) \Big]^0
+\ldots \bigg)T_i^+
-
\\
-
(N^0+N^1+\ldots) e^{\overline x_i}
\ +
\end{multline*}

\begin{multline*}
+ \frac{1}{2m+1} \sum_{i=1}^{n-1}
\bigg(
N^0+N^1-
\Big[ N\Big(\sum_{j\ne i}   \left(
\frac{2l+1}{-\overline k_i+\overline k_j} + \frac{2l+1}{-\overline k_i-\overline
k_j}\right)
-\frac{(2m+1)l}{\overline k_i}
+
\\
+
\frac{2m+1}{-\overline k_i+\overline k_n-l+m}+
\frac{2m+1}{-\overline k_i-\overline k_n-l+m} \Big)\Big]^0+\ldots
\bigg)T_i^-
-
\\
-(N^0+N^1+\ldots) e^{-\overline x_i}\ +
\end{multline*}

\begin{multline*}
+ \frac{1}{2l+1}
\bigg(
N^0+N^1-
\Big[    \Big(
\frac{(2l+1)m}{\overline k_n}
+
\\
+
\sum_{j=1}^{n-1}\Bigl(
\frac{2l+1}{\overline k_n+\overline k_j+l-m}+
\frac{2l+1}{\overline k_n-\overline k_j+l-m}\Bigr)
\Big) N\Big]^0+\ldots
\bigg)T_n^+
-
\\
-(N^0+N^1+\ldots) e^{\overline x_n}
\ +
\end{multline*}

\begin{multline*}
+ \frac{1}{2l+1}
\bigg(
N^0+N^1-
\Big[    \Big(
\frac{(2l+1)m}{-\overline k_n}
+
\\
+
\sum_{j=1}^{n-1}\Bigl(
\frac{2l+1}{-\overline k_n+\overline k_j+l-m}+
\frac{2l+1}{-\overline k_n-\overline k_j+l-m}\Bigr)
\Big) N\Big]^0+\ldots
\bigg)T_n^-
-
\\
-(N^0+N^1+\ldots) e^{-\overline x_n}
\Biggr\}
\ \times
\end{multline*}
$$
\times
\left(P_s^0+P_s^1+\ldots\right) e^{(k,x)}.
$$
Applying operators $T_i^\pm$ we get the following expression
\begin{multline*}
\frac{1}{2m+1} \sum_{i=1}^{n-1}
\Biggl\{
\biggl[\biggl( -\sum_{j\ne i} \Bigl(
\frac{2l+1}{\overline k_i+\overline k_j}+
\frac{2l+1}{\overline k_i-\overline k_j}\Bigr)-
\\
-
\frac{2m+1}{\overline k_i+\overline k_n-l+m}-
\frac{2m+1}{\overline k_i-\overline k_n-l+m}
-\frac{(2m+1)l}{\overline k_i}
\biggr)N \biggr]^0 e^{\overline x_i} P_s^0
+
\\
+
e^{\overline x_i} N^0 \frac{\p P_s^0}{\p \overline k_i} (2m+1)
+\ldots
\Biggr\} \frac{e^{(k,x)}}{N}+
\end{multline*}
\begin{multline*}
+
\frac{1}{2m+1} \sum_{i=1}^{n-1}
\Biggl\{
\biggl[ \biggl(-\sum_{j\ne i} \Bigl(
\frac{2l+1}{-\overline k_i+\overline k_j}+
\frac{2l+1}{-\overline k_i-\overline k_j}\Bigr)-
\\
-
\frac{2m+1}{-\overline k_i+\overline k_n-l+m}-
\frac{2m+1}{-\overline k_i-\overline k_n-l+m}
+\frac{(2m+1)l}{\overline k_i}
\biggr)N \biggr]^0 e^{-\overline x_i} P_s^0
-
\\
-
e^{-\overline x_i} N^0 \frac{\p P_s^0}{\p \overline k_i} (2m+1)
+\ldots
\Biggr\} \frac{e^{(k,x)}}{N}+
\end{multline*}
\begin{multline*}
+
\frac{1}{2l+1}
\Biggl\{
\biggl[\biggr(
-\sum_{j=1}^{n-1} \Bigl(
\frac{2l+1}{\overline k_n+\overline k_j+l-m}+
\frac{2l+1}{\overline k_n-\overline k_j+l-m}\Bigr)
-
\\-
\frac{(2l+1)m}{\overline k_n}
\biggr)N \biggr]^0 e^{\overline x_n} P_s^0
+
e^{\overline x_n} N^0 \frac{\p P_s^0}{\p \overline k_n} (2l+1)
+\ldots
\Biggr\} \frac{e^{(k,x)}}{N}+
\end{multline*}
\begin{multline*}
+
\frac{1}{2l+1}
\Biggl\{
\biggl[\biggr(
-\sum_{j=1}^{n-1} \Bigl(
\frac{2l+1}{-\overline k_n+\overline k_j+l-m}+
\frac{2l+1}{-\overline k_n-\overline k_j+l-m}\Bigr)
+
\\+
\frac{(2l+1)m}{\overline k_n}
\biggr)N \biggr]^0 e^{-\overline x_n} P_s^0
-
e^{\overline x_n} N^0 \frac{\p P_s^0}{\p \overline k_n}
(2l+1)+\ldots
\Biggr\} \frac{e^{(k,x)}}{N}.
\end{multline*}

We assume now that $P_s^0$ has the following form
$$
P_s^0=\sum_{\{\lambda\}} c_\lambda P_{s,\{\lambda\}}^0,
$$
where
$$
P_{s,\{\lambda\}}^0= \prod_{i<j}^n
\overline k_{j}^{\lambda_j}\,
(\overline k_i+\overline k_j)^{\lambda_{ij}^+}\,
(\overline k_i-\overline k_j)^{\lambda_{ij}^-}.
$$
Then $P_{s+1}^0$ being the ratio of the highest term in the numerator to
the highest term in the denominator takes the following form
$$
P_{s+1}^0=\sum_{\{\lambda\}} c_\lambda P_{s+1,\{\lambda\}}^0,
$$
where
\begin{multline*}
P_{s+1,\{\lambda\}}^0=
\sum_{i=1}^{n-1}
(e^{\overline x_i}-e^{-\overline x_i})
\biggl(
\frac{\lambda_i-l}{\overline k_i}+
\sum_{j\ne i}^{n-1} \Bigl(
\frac{\lambda_{ij}^{+} - \frac{2l+1}{2m+1}}{\overline k_i+\overline k_j} +
\frac{\lambda_{ij}^{-} - \frac{2l+1}{2m+1}}{\overline k_i-\overline k_j}
\Bigr) +
\\
\null\hfill
+
\frac{\lambda_{in}^{+} -1}{\overline k_i+\overline k_n}+
\frac{\lambda_{in}^{-} -1}{\overline k_i-\overline k_n}
\biggr)
P_{s,\{\lambda\}}^0+
\\[2mm]
+
(e^{\overline x_n}-e^{-\overline x_n})
\biggl(
\frac{\lambda_n-m}{\overline k_n}+
\sum_{j\ne i}^{n-1} \Bigl(
\frac{\lambda_{jn}^{+} - 1}{\overline k_j+\overline k_n} +
\frac{-\lambda_{jn}^{-}  +1}{\overline k_j-\overline k_n}
\Bigr)
\biggr)
P_{s,\{\lambda\}}^0
\end{multline*}
and we assume the notations $\lambda_{ij}^\pm=\lambda_{ji}^\pm$.
Thus finally we have
\begin{multline*}
P_{s+1, \{\lambda\}}^0=
\sum_{{i_0}=1}^{n-1}
(\lambda_{i_0}-l)(e^{\overline x_{i_0}}-e^{-\overline x_{i_0}})
\overline k_{i_0}^{\lambda_{i_0}-1}
\prod_{j\ne {i_0}} \overline k_j^{\lambda_j}
\prod_{i<j}^n
(\overline k_i+\overline k_j)^{\lambda_{ij}^+}
(\overline k_i-\overline k_j)^{\lambda_{ij}^-}
+
\\
+
(\lambda_n-m)(e^{\overline x_n}-e^{-\overline x_n})
\overline k_n^{\lambda_n-1}
\prod_{j\ne n} \overline k_j^{\lambda_j}
\prod_{i<j}^n
(\overline k_i+\overline k_j)^{\lambda_{ij}^+}
(\overline k_i-\overline k_j)^{\lambda_{ij}^-}
\ +
\end{multline*}
\begin{multline*}
+
\sum_{{i_0}<{j_0}}^{n-1}
\Bigl(\lambda_{{i_0}{j_0}}^{+}-\frac{2l+1}{2m+1}\Bigr)
(e^{\overline x_{i_0}}-e^{-\overline x_{i_0}}+e^{\overline x_{j_0}}-e^{-\overline x_{j_0}})
(\overline k_{i_0}+\overline k_{j_0})^{\lambda_{{i_0}{j_0}}^{+}-1}
\times
\\
\times
\prod_{j=1}^n        \overline k_j^{\lambda_j}
\prod_{\nad{i<j}{(i,j)\ne ({i_0},{j_0})}}
(\overline k_i+\overline k_j)^{\lambda_{ij}^+}
\prod_{\nad{i<j}{(i,j)\ne ({i_0},{j_0})}}
(\overline k_i-\overline k_j)^{\lambda_{ij}^-}
\ +
\end{multline*}
\begin{multline*}
+
\sum_{{i_0}<{j_0}}^{n-1}
\Bigl(\lambda_{{i_0}{j_0}}^{-}-\frac{2l+1}{2m+1}\Bigr)
(e^{\overline x_{i_0}}-e^{-\overline x_{i_0}}-e^{\overline x_{j_0}}+e^{-\overline x_{j_0}})
(\overline k_{i_0}-\overline k_{j_0})^{\lambda_{{i_0}{j_0}}^{-}-1}
\times
\\
\times
\prod_{j=1}^n         \overline k_j^{\lambda_j}
\prod_{\nad{i<j}{(i,j)\ne ({i_0},{j_0})}}
(\overline k_i+\overline k_j)^{\lambda_{ij}^+}
\prod_{\nad{i<j}{(i,j)\ne ({i_0},{j_0})}}
(\overline k_i-\overline k_j)^{\lambda_{ij}^-}
\ +
\end{multline*}
\begin{multline*}
+
\sum_{{i_0}=1}^{n-1}
(\lambda_{{i_0}n}^{+}-1)
(e^{\overline x_{i_0}}-e^{-\overline x_{i_0}}+e^{\overline x_n}-e^{-\overline x_n})
(\overline k_{i_0}+\overline k_n)^{\lambda_{{i_0}n}^{+}-1}
\times
\\
\times
\prod_{j=1}^n         \overline k_j^{\lambda_j}
\prod_{\nad{i<j}{(i,j)\ne ({i_0},n)}}
(\overline k_i+\overline k_j)^{\lambda_{ij}^+}
\prod_{\nad{i<j}{(i,j)\ne ({i_0},n)}}
(\overline k_i-\overline k_j)^{\lambda_{ij}^-}
\,+
\end{multline*}
\begin{multline}
+
\sum_{{i_0}=1}^{n-1}
(\lambda_{{i_0}n}^{-}-1)
(e^{\overline x_{i_0}}-e^{-\overline x_{i_0}}-e^{\overline x_n}+e^{-\overline x_n})
(\overline k_{i_0}-\overline k_n)^{\lambda_{{i_0}n}^{-}-1}
\times
\\
\times
\prod_{j=1}^n         \overline k_j^{\lambda_j}
\prod_{\nad{i<j}{(i,j)\ne ({i_0},n)}}
(\overline k_i+\overline k_j)^{\lambda_{ij}^+}
\prod_{\nad{i<j}{(i,j)\ne ({i_0},n)}}
(\overline k_i-\overline k_j)^{\lambda_{ij}^-}.
\label{ps+1}
\end{multline}

Let us now follow how  $P^0$ is changing starting from
$$
\varphi_0=
\prod_{\alpha\in \cnm}  \prod_{j=1}^{m_\alpha}
(k+j\alpha,\alpha) (k-j\alpha,\alpha) e^{(k,x)},
$$
that is
$$
P_0^0=
\prod_{i=1}^n  \overline k_i^{2m_i}
\prod_{i<j}^n   (\overline k_i+\overline k_j)^{2m_{ij}}
\prod_{i<j}^n   (\overline k_i-\overline k_j)^{2m_{ij}}.
$$
Formula (\ref{ps+1}) shows that for anyÿ $s$ \ $P_s^0$ is a
linear combination of monomials consisting of the products
$\overline k_i^{\lambda_i}  (\overline k_i+\overline k_j)^{\lambda_{ij}^+}
(\overline k_i-\overline k_j)^{\lambda_{ij}^-}$, and the degree of
monomials is decreasing by 1 at every application of the operator
$D$. Besides this the coefficients in formula (\ref{ps+1}) show
that monomials with degrees $\lambda_i<m_i$ and $\lambda_{ij}^\pm
< m_{ij}$ cannot appear. Thus we get
$$
P_{_{\sum m_\alpha}}^0 = c(x) \prod_{i=1}^n \overline k_i^{m_i}
\prod_{i<j}   (\overline k_i+\overline k_j)^{m_{ij}}
(\overline k_i-\overline k_n)^{m_{ij}}.
$$
Therefore function $c(x)^{-1} \varphi_{_{\sum m_\alpha}}$ satisfies conditions
(\ref{psi12}), (\ref{22}) of the BA function.

We are left to determine the coefficient $c(x)$. For this we
analyze once again formula (\ref{ps+1}). At every step one of the
terms
 $\overline k_i$, $\overline k_i\pm \overline k_j$ in the monomials is changed
 by the corresponding function of $x$ with some coefficient. We begin with
 the monomial
$$
\prod_{j=1}^n \overline k_j^{2m_j}
\prod_{i<j}   (\overline k_i+\overline k_j)^{2m_{ij}}
        (\overline k_i-\overline k_j)^{2m_{ij}}
$$
and finish by the monomial
$$
\prod_{j=1}^n \overline k_j^{m_j}
\prod_{i<j}   (\overline k_i+\overline k_j)^{m_{ij}}
        (\overline k_i-\overline k_j)^{m_{ij}}.
$$
Therefore
\begin{multline*}
c(x)=c_0 \prod_{i=1}^{n-1} (e^{\overline x_i}-e^{-\overline x_i})^l
(e^{\overline x_n}-e^{-\overline x_n})^m
\times
\\
\times
\prod_{i<j}^{n-1}
(e^{\overline x_i}-e^{-\overline x_i}+e^{\overline x_j}-e^{-\overline x_j})^\frac{2l+1}{2m+1}
\prod_{i<j}^{n-1}
(e^{\overline x_i}-e^{-\overline x_i}-e^{\overline x_j}+e^{-\overline x_j})^\frac{2l+1}{2m+1}
\times
\\
\null\hfill\times
\prod_{i=1}^{n-1}
(e^{\overline x_i}-e^{-\overline x_i}+e^{\overline x_n}-e^{-\overline x_n})
(e^{\overline x_i}-e^{-\overline x_i}-e^{\overline x_n}+e^{-\overline x_n})
=
\\
=
c_0 \,(e^{\overline x_n}-e^{-\overline x_n})^m
\times
\\
\times
\prod_{i=1}^{n-1}
(e^{\overline x_i}-e^{-\overline x_i})^l
\prod_{i<j}^{n-1}
(e^{\overline x_i-\overline x_j}-e^{\overline x_j-\overline x_i})^\frac{2l+1}{2m+1}
(e^{\overline x_i+\overline x_j}-e^{-\overline x_i-\overline x_j})^\frac{2l+1}{2m+1}
\times\hfill\null
\\
\times
\prod_{i=1}^{n-1}
(e^{\overline x_i-\overline x_n}-e^{\overline x_n-\overline x_i})
(e^{\overline x_i+\overline x_n}-e^{-\overline x_i-\overline x_n}).
\end{multline*}
It is left to determine the coefficient $c_0$. This is an integer
equal to the total number of possible monomials. From (\ref{ps+1})
it easily follows that at the first step there appear $M=\sum m_i
+2\sum m_{ij}$ monomials, and after the second step there appears
$M(M-1)$ monomials. In total we obtain $c_0=M!$ and theorem is
proven.
\end{proof}

In the end of this section we put the result on bispectrality.

\begin{theorem}\lb{bispA2}
The Baker--Akhiezer function $\psi(k,x)$ for the system $\cnm$ satisfies the following equation
in variables $k$:
$$
D \psi(k,x) = \left(\frac{2}{2m+1}\sum_{j=1}^{n-1} \cosh \sqrt{2m+1} \, x_j +
\frac{2}{2l+1} \cosh \sqrt{2l+1} \, x_n \right) \psi (k,x),
$$
where $D$ is operator \mref{8}.
For any polynomial $p(k)\in R_{\cnm}$
the difference operators
$$
D_p = ad_D^{\deg p} p(k)
$$
commute. These operators also commute with operator $D$.

\end{theorem}

\begin{proof}
 In the notations  \mref{aa}, \mref{bb} from theorem \ref{teorCsyst} and
 propositions \ref{predl4}, \ref{predl5} it follows that $\varphi_{_{\sum m_\alpha}+1}$ has the form $P(k,x) e^{(k,x)}$ where $P$ is a polynomial in
$k$ of degree less than $\sum m_\alpha$, and it satisfies axiomatics \mref{22}.
By lemma \ref{lem1} it follows that
$\varphi_{_{\sum m_\alpha}+1}=0$
which is equivalent to the first statement of the theorem.

Now, as it is explained in section \ref{gl2sect1} for any $p\in R_{\cnm}$ 
there exists differential operator $L_p(x,\p_x)$ such that
$$
L_p(x,\p_x) \psi(k,x)=p(k)\psi(k,x).
$$
By bispectrality arguments presented in section \ref{gl2sect2} we have
$$
D_p \psi (k,x) = a_p(x) \psi(k,x)
$$
for some function $a_p(x)$, therefore we have the relation
$$
(D_{p_1} D_{p_2}-D_{p_2} D_{p_1}) \psi(k,x) =
(a_{p_1}a_{p_2}-a_{p_2}a_{p_1})\psi(k,x)=0.
$$
Because of special form of $\psi$ it follows that $D_{p_1} D_{p_2}-D_{p_2}D_{p_1}
=0$.
\end{proof}

\section{Configuration  $\pmb A_{n,2}(m)$}

The vectors and multiplicities forming this system in $\C^{n+1}$ are as follows.
The vectors
$\alpha_{0i}=\sqrt{-m-1}
e_0-e_i$, $e_i-\sqrt{m}e_n$ have multiplicities $m_{0i}=m_{in}=1$, $i=1,\ldots,n-1$.
The vectors $\alpha_{ij}= e_i-e_j$ have multiplicities $m_{ij}=m$, $1\le i<j\le n-1$, the vector $\alpha_{0n}=
\sqrt{-m-1}\,e_0 - \sqrt{m}\,e_n$  has multiplicity $m_{0n}=1$.

This configuration was introduced by Chalykh and Veselov in
\cite{CV2} as the one satisfying the rational locus conditions but
not satisfying the $\vee$-conditions and thus not leading to a
solution of generalized WDVV equations (see \cite{CV2}). In the
case $n=2$ the system contains three vectors all having multiplicity $1$
thus parameter $m$ can be arbitrary complex rather than
integer. The corresponding elliptic operator was considered by
Hietarinta \cite{H} (see also \cite{CEO}, \cite{CFV3}). The
important for us feature of this configuration is the fact that
the system does not admit the Baker--Akhiezer function in the
sense of \cite{VSCh}, that is satisfying the conditions
$$
\psi(k+s \a,x) = \psi (k-s \a, x)
$$
at $(\a,k)=0$, $s\le m_\a$. But the system admits BA function in the sense
of our definition that is we impose conditions  \mref{22}.

In order to construct BA function we again follow the scheme of
\cite{ChalBisp}. As difference operator $D$ we take
    \begin{equation}
    \label{DAnmC}
    D=\sum_{i=0}^n \frac1{{\overline e_i}^2} \prod_{\nad{j=0}{j\ne i}}^n
    (k-m_{ij}\alpha_{ij}, \alpha_{ij}) T_i
\frac1{\prod_{\nad{j=0}{j\ne i}}^n
    (k-\alpha_{ij}, \alpha_{ij})},
\end{equation}
where for this section we have introduced the following notations
$$
\overline e_0 = \sqrt{-m-1}\, e_0, \,
\overline e_n =\sqrt{m} e_n, \quad
\mbox{ and } \overline e_i = e_i \,
\mbox{ forÿ} \,\,  1\le i\le{n-1}.
$$
Also for $0\le i \le n$ we  denote
$$
{\overline e_i}^2 = (\overline e_i,\overline e_i), \,  \overline k_i = (k, \overline e_i), \, \overline x_i = (x, \overline e_i)
$$
for this section.
Operators $T_i$ act by the rule $T_i(f(k)) =
f(k+2\overline e_i)$, and we understand that
$\a_{ij}=\overline e_i - \overline e_j$  also when $i>j$. Then operator $D$ can be written as follows
\begin{multline*}
D=-\frac1{m+1}
\left(1+\frac{2(m+1)}{\overline k_0-\overline k_n -2m-1} \right)\prod_{j=1}^{n-1}
\left(1+\frac{2(m+1)}{\overline k_0-k_j-m-2}\right)T_0+
\\
\sum_{i=1}^{n-1}
\left(1-\frac{2}{k_i-\overline k_0+m+2}\right)\left(1-\frac{2}{k_i-\overline k_n-m+1}\right)
\prod_{j=1}^{n-1} \left(1-\frac{2m}{k_i-k_j}\right) T_i+
\\
+\frac1{m}
\left(1-\frac{2m}{\overline k_n- \overline k_0+2m+1}\right)\prod_{j=1}^{n-1}
\left(1-\frac{2m}{\overline k_n-k_j+m-1}\right) T_n.
\end{multline*}
At first we rearrange conditions (\ref{2'}) into more convenient for us form
similarly to the case $C_n(l,m)$ system considered earlier. Namely, for $\a=e_i-e_j$, $1\le i<j\le n-1$ dropping $A_+$ in the notation $\psi_\a^{A_+}(k)$ for simplicity,
 the condition
$\psi_\a(k+s\a)=\psi_\a(k-s\a)$ is equivalent to condition
\beq{aksA0}
T^s_i\psi_\a = T^s_j\psi_\a \quad \mbox{at  } \ k\in\pi_{ij}:
k_i-k_j=0.
\eeq
Now we move to consideration of the condition for $\a=\overline e_i - \overline
e_j$, where  $i=0$ or $j=n$ or both. ÒThe identity $\psi_\a(k+\a)=\psi_\a(k-\a)$
is equivalent to condition
\beq{aksA}
T_i\psi_\a = T_j\psi_\a  \quad \mbox{at  } \ k\in\pi_{ij}:
\overline k_i-\overline k_j +\overline e_i^2-\overline e_j^2=0,
\eeq
where $\overline k_i = (k,\overline e_i)$. Indeed, let
$k\in\pi_{ij}$, then
$$
T_i\psi_\a(k)=\psi_\a(k+2\overline e_i)=\psi_\a(k+\a+\overline e_i+\overline
e_j),
$$
$$
T_j\psi_\a(k)=\psi_\a(k+2\overline e_j)=\psi_\a(k-\a+\overline e_i+\overline
e_j).
$$
As $(\a,k+\overline e_i+\overline e_j)=0$ the conditions in the original form are equivalent
to (\ref{aksA}).

\begin{predl}\label{predl6}
For any holomorphic function $f(k_0,k_1,\ldots,k_n)$ satisfying
conditions $(\ref{aksA0}),$ $(\ref{aksA})$ the function
$Df(k_0,\ldots,k_n)$ is also holomorphic if $D$ is given by
$(\ref{DAnmC})$.
\end{predl}
\begin{proof}
In principle function $Df(k_0,\ldots,k_n)$ could have
singularities at the hyperplane $\pi$ of the form \beq{pi}
T_{i_0}(k-\alpha_{i_0 j_0},\alpha_{i_0 j_0})=(k+\overline e_{i_0} +\overline
e_{j_0}, \overline e_{i_0} - \overline e_{j_0}) = 0, \eeq $i_0\ne j_0$. We
will show that this does not happen. We collect terms in $Df$
which possibly have singularities at $(k+\overline e_{i_0}+\overline
e_{j_0},\overline e_{i_0}-\overline e_{j_0}) = 0$. Since
$$
T_{i_0}(k-\alpha_{i_0 j_0},\alpha_{i_0 j_0}) =
- T_{j_0}(k-\alpha_{j_0 i_0},\alpha_{j_0 i_0}) =
(k+\overline e_{i_0} +\overline e_{j_0},
\overline e_{i_0} - \overline e_{j_0}),
$$
we get sum of two terms
\begin{multline}\label{dveosob}
\frac1{
(k+\overline e_{i_0} +\overline e_{j_0},
\overline e_{i_0} - \overline e_{j_0})}
\Bigg(
\frac1{\overline e_{i_0}^2} \prod_{\nad{j=0}{j\ne i_0}}^n
    (k-m_{i_0 j}, \alpha_{i_0 j}) T_{i_0}
\frac{f(k)}{\prod_{j\ne i_0, j_0}^n
    (k-\alpha_{i_0 j}, \alpha_{i_0 j})} -
\\-
\frac1{\overline e_{j_0}^2} \prod_{\nad{i=0}{i\ne j_0}}^n
    (k-m_{j_0 i}, \alpha_{j_0 i}) T_{j_0}
\frac{f(k)}{\prod_{i\ne i_0, j_0}^n
    (k-\alpha_{j_0 i}, \alpha_{j_0 i})}
\Bigg).
\end{multline}
We have to show that the expression in brackets vanishes at
$k\in\pi$ (\ref{pi}). We note that the vectors $A_{i_0 j_0} =
\{\alpha_{i_0 j}, \alpha_{j_0 i}|\ j\ne i_0, i\ne j_0, i_0\}$ lie
in some half-space in $\mathbb C^n \approx \mathbb R^{2n}$, and
for any choice of the subsystem $B\subset A$ such that $B\cup
A_{i_0 j_0}$ is a positive system $A_{+}$, the vector $\alpha_{i_0
j_0}$ is edge vector in $A_+$. Therefore axiomatic conditions
(\ref{aksA0}), (\ref{aksA}) state, in particular, that
{\normalsize
\begin{multline}\label{gm}
T_{i_0}\frac{f(k)}{\prod_{j\ne i_0, j_0} \prod_{s=1}^{m_{i_0 j}}
(k-s \alpha_{i_0 j},\alpha_{i_0 j})
\prod_{i\ne i_0, j_0}\prod_{s=1}^{m_{j_0 i}}
(k-s\alpha_{j_0 i}, \alpha_{j_0 i})\prod_{\beta\in B}\prod_{s=1}^{m_\b} (k-s\beta,\beta)}
=
\\
=
T_{j_0}\frac{f(k)}{\prod_{i\ne i_0, j_0}^n \prod_{s=1}^{m_{j_0 i}}
(k-s \alpha_{j_0 i},\alpha_{j_0 i})
\prod_{i\ne i_0, j_0}\prod_{s=1}^{m_{i_0 j}}
(k-s\alpha_{i_0 j}, \alpha_{i_0 j})\prod_{\beta\in B}\prod_s (k-s\beta,\beta)}
\end{multline}

}
\noindent
on the hyperplane $\pi$. Further we apply shift operators to a part of the
product in (\ref{gm}) and we use equality
$$
T_{i_0} \prod_{s=2}^{m_{i_0 j}}
(k-s \alpha_{i_0 j},\alpha_{i_0 j}) =
\prod_{s=1}^{m_{i_0 j}-1}
(k-s \alpha_{i_0 j},\alpha_{i_0 j}),
$$
which is non-trivial only if $m_{i_0 j}>1$, that is for
$1\le i_0,j\le n-1$.
We get
\begin{multline*}
\frac{T_{i_0}\frac{f(k)}{\prod_{j\ne i_0,j_0} (k- \alpha_{i_0 j},\alpha_{i_0 j})}}
{\prod_{j\ne i_0, j_0} \prod_{s=1}^{m_{i_0 j}-1}
(k-s \alpha_{i_0 j},\alpha_{i_0 j})
\prod_{i\ne i_0, j_0}\prod_{s=1}^{m_{j_0 i}}
(k-s\alpha_{j_0 i}, \alpha_{j_0 i})}
=\\
=
\frac{T_{j_0}\frac{f(k)}{\prod_{i\ne i_0,j_0}
(k-\alpha_{j_0 i}, \alpha_{j_0 i})}}{\prod_{i\ne i_0, j_0}^n \prod_{s=1}^{m_{j_0 i}-1}
(k-s \alpha_{j_0 i},\alpha_{j_0 i})
\prod_{j\ne i_0, j_0}\prod_{s=1}^{m_{i_0 j}}
(k-s\alpha_{i_0 j}, \alpha_{i_0 j})}.
\end{multline*}
After necessary cancellations we obtain from above
\begin{multline*}
\prod_{j\ne i_0, j_0} (k-m_{i_0 j}\alpha_{i_0 j}, \alpha_{i_0 j})
T_{i_0}\frac{f(k)}{\prod_{j\ne i_0,j_0} (k- \alpha_{i_0 j},\alpha_{i_0 j})}
=
\\=
\prod_{i\ne i_0, j_0} (k-m_{j_0 i}\alpha_{j_0 i} , \alpha_{j_0 i})
T_{j_0}\frac{f(k)}{\prod_{i\ne i_0,j_0} (k- \alpha_{j_0 i},\alpha_{j_0
i})}.
\end{multline*}
Simplifying (\ref{dveosob}) with the help of equality
$$
\frac{(k-m_{i_0 j_0}\alpha_{i_0 j_0}, \alpha_{i_0 j_0})}
{(\overline e_{i_0}, \overline e_{i_0})} =
\frac{(k-m_{j_0 i_0}\alpha_{j_0 i_0}, \alpha_{j_0 i_0})}
{(\overline e_{j_0}, \overline e_{j_0})}
$$
which is valid for $k\in\pi$, we conclude that expression (\ref{dveosob})
has no singularities at the hyperplane $\pi$.
\end{proof}

\begin{predl}\label{predl7} Let holomorphic function $f(k)$ satisfy conditions
$(\ref{aksA0}),$ $(\ref{aksA})$. Then the function
$Df(k)$ also satisfies $(\ref{aksA0}),$ $(\ref{aksA})$ if $D$ is given by
$(\ref{DAnmC})$.
\end{predl}

Before we start proving the proposition we state a lemma which will be useful
for us to work with axiomatic conditions (\ref{aksA0}), (\ref{aksA}).

\begin{lemma}\lb{lm23}
Let vector $\alpha_{ij}\in  A_{n,2}(m)$ be edge vector for two
subsystems $A_+^{(1)}$ and $A_+^{(2)}$. Then the following
condition for the holomorphic function $f(k)$
$$
(T_i^s - T_j^s)\frac{f(k)}{\prod_{\nad{\beta\in A_+^{(1)}}{\beta\ne\alpha_{ij}}}
\vec \beta} =0  \qquad  \mbox{at } \ k\in\pi_{ij}, \ 1\le s \le
m_{ij}
$$
is equivalent to the condition
$$
(T_i^s - T_j^s)\frac{f(k)}{\prod_{\nad{\beta\in A_+^{(2)}}{\beta\ne\alpha_{ij}}}
\vec \beta} =0 \qquad   \mbox{at } \  k\in\pi_{ij}, \ 1\le s
\le m_{ij}
$$
where $\vec\b=\prod_{l=1}^{m_\b}(k+l\b,\b)$.
\end{lemma}
\begin{proof}
We denote for the brevity $\prod_t \vec\beta =
\prod_{\nad{\beta\in A_+^{(t)}}{\beta\ne\alpha_{ij}}}
\vec \beta$, \hbox{$t=1,2$}. As
$$
(T_i^s-T_j^s)\frac{f(k)}{\prod_2\vec\beta} =
\bigg(\frac{T_i^s f(k)}{T_i^s \prod_1\vec\beta}\bigg)
\bigg( \frac{T_i^s \prod_1\vec\beta}{T_i^s \prod_2\vec\beta}\bigg)
 -
\bigg(\frac{T_j^s f(k)}{T_j^s \prod_1\vec\beta}\bigg)
\bigg( \frac{T_j^s \prod_1\vec\beta}{T_j^s \prod_2\vec\beta}\bigg),
$$
we have to show that
$$
T_i^s \frac{\prod_1\vec\beta}{\prod_2\vec\beta} =
T_j^s \frac{\prod_1\vec\beta}{\prod_2\vec\beta} \qquad \mbox{at }  \ k\in\pi_{ij}.
$$
This is equivalent to
\beq{Tij}
(T_i^s-T_j^s) \frac{\prod_{\nad{\beta \in A_+^{(1)}}{(\beta,\alpha_{ij})\ne 0}}
\vec\beta}
{\prod_{\nad{\beta \in A_+^{(2)}}{(\beta,\alpha_{ij})\ne 0}}\vec\beta}
=0\qquad \mbox {at }   k\in\pi_{ij}.
\eeq
Regrouping the product terms condition (\ref{Tij}) takes the form
$$
(T_i^s-T_j^s) \prod_{q\ne i,j}\prod_{t=1}^{m_{jq}}\prod_{s=1}^{m_{iq}}
\frac{(k+s\varepsilon_{iq}\alpha_{iq},\alpha_{iq})
(k+t\varepsilon_{jq}\alpha_{jq},\alpha_{jq})}
{(k+s\delta_{iq}\alpha_{iq},\alpha_{iq})
(k+t\delta_{jq}\alpha_{jq},\alpha_{jq})} =0
 \qquad   \mbox {at } k\in\pi_{ij},
$$
where $\varepsilon_{iq},\delta_{iq}=\pm 1$. And it is sufficient to show
that $\forall q\ne i,j$
\beq{Tij2}
(T_i^s-T_j^s) \prod_{t=1}^{m_{jq}}\prod_{s=1}^{m_{iq}}
\frac{(k+s\varepsilon_{iq}\alpha_{iq},\alpha_{iq})
(k+t\varepsilon_{jq}\alpha_{jq},\alpha_{jq})}
{(k+s\delta_{iq}\alpha_{iq},\alpha_{iq})
(k+t\delta_{jq}\alpha_{jq},\alpha_{jq})} =0
 \qquad   \mbox {at } k\in\pi_{ij}.
\eeq
This means condition (\ref{Tij}) is reduced to the
two-dimensional identity (\ref{Tij2}) in the plane containing
vectors $\alpha_{ij},\alpha_{iq},\alpha_{jq}$. And the condition
that $\alpha_{ij}$ is edge vector for $A_+^{(1)}$,
 $A_+^{(2)}$ means that
$\alpha_{ij}=\pm(\varepsilon_{iq}\alpha_{iq}-\varepsilon_{jq}\alpha_{jq})$,
that is  $\varepsilon_{iq}=\varepsilon_{jq}$, analogously we have $\delta_{iq}=\delta_{jq}$.
Therefore property (\ref{Tij2}) is reduced to the identity
\beq{Tij3}
(T_i^s-T_j^s) \prod_{t=1}^{m_{jq}}\prod_{s=1}^{m_{iq}}
\frac{(k+s\alpha_{iq},\alpha_{iq})
(k+t\alpha_{jq},\alpha_{jq})}
{(k-s\alpha_{iq},\alpha_{iq})
(k-t\alpha_{jq},\alpha_{jq})} =0
 \qquad   \mbox {at } k\in\pi_{ij}.
\eeq
Now we separately consider the arising cases

a) $1\le i,j\le n-1$. At any $q$ the product in (\ref{Tij3}) is
invariant under the reflection $k_i \leftrightarrow k_j$.
Therefore in particular property~(\ref{Tij3})holds.

Further we may assume that  $s=1$.

b) $i=0$, $1\le j\le n-1$. Consider firstly the case $q<n$. Identity~(\ref{Tij3})
takes the form
$$
(T_0-T_j)
\frac{(k+\alpha_{0q},\alpha_{0q})}
{(k-\alpha_{0q},\alpha_{0q})}
\frac{\prod_{t=1}^m (k+t\alpha_{jq},\alpha_{jq})}
{\prod_{t=1}^m (k-t\alpha_{jq},\alpha_{jq})} =0
$$
at $\overline k_0 - \overline k_j +(\overline e_0,\overline e_0) - (\overline e_j,\overline e_j) =0$. Or,
more explicitly, we have
\begin{multline*}
\frac{(k+2e_0+\alpha_{0q},\alpha_{0q})}
{(k+2e_0-\alpha_{0q},\alpha_{0q})}
\frac{\prod_{t=1}^m (k_j-k_q+2t)}
{\prod_{t=1}^m (k_j-k_q-2t)} =
\\=
\frac{(k+\alpha_{0q},\alpha_{0q})}
{(k-\alpha_{0q},\alpha_{0q})}
\frac{\prod_{t=1}^m (k_j-k_q+2t+2)}
{\prod_{t=1}^m (k_j-k_q-2t+2)}.
\end{multline*}
Performing the cancellations recalling that  $\overline k_0 = \overline k_j - (\overline e_0,\overline e_0) +
(\overline e_j,\overline e_j) =0$ we get
$$
\frac{(k_j-k_q-2m)(k_j-k_q+2)}{(k_j-k_q)(k_j-k_q-2m)}
=\frac{(k_j-k_q+2)(k_j-k_q+2m+2)}{(k_j-k_q+2(m+1))(k_j-k_q)},
$$
which is obviously satisfied. Further we consider relation
(\ref{Tij3}) at $q=n$. We have to check that
$$
T_0 \frac{(k+\alpha_{0n},\alpha_{0n})
(k+\alpha_{jn},\alpha_{jn})}
{(k-\alpha_{0n},\alpha_{0n})
(k-\alpha_{jn},\alpha_{jn})} =
T_j \frac{(k+\alpha_{0n},\alpha_{0n})
(k+\alpha_{jn},\alpha_{jn})}
{(k-\alpha_{0n},\alpha_{0n})
(k-\alpha_{jn},\alpha_{jn})}
$$
at $\overline k_0 - \overline k_j +(\overline e_0,\overline e_0) - (\overline e_j,\overline e_j) =0$.
Applying difference operators we have
\begin{multline*}
\frac{(\overline k_0-\overline k_n+3\overline e_0^2+\overline e_n^2)(\overline k_j-\overline k_n+\overline e_j^2+\overline e_n^2)}
{(\overline k_0-\overline k_n+\overline e_0^2-\overline e_n^2)
(\overline k_j-\overline k_n-\overline e_j^2-\overline e_n^2)}
=
\\=
\frac{(\overline k_0-\overline k_n+\overline e_0^2+\overline e_n^2)(\overline k_j-\overline k_n+3\overline e_j^2+\overline e_n^2)}
{(\overline k_0-\overline k_n-\overline e_0^2-\overline e_n^2)(\overline k_j-\overline k_n+\overline e_j^2-\overline e_n^2)}.
\end{multline*}
We substitute now $\overline k_0 = \overline k_j - \overline e_0^2 +\overline e_j^2$,
$\overline e_0^2= -m-1$, $\overline e_n^2=m$, $\overline e_j^2=1$ and we get obvious identity
\begin{multline*}
\frac{(\overline k_j-\overline k_n-m-1)(\overline k_j-\overline k_n+m+1)}{(\overline k_j-\overline k_n-m+1)(\overline k_j-\overline k_n-m-1)}
=
\\=
\frac{(\overline k_j-\overline k_n+m+1)(\overline k_j-\overline k_n+m+3)}{(\overline k_j-\overline k_n+m+3))(\overline k_j-\overline k_n-m+1)}.
\end{multline*}

c) $i=0$, $j=n$. We have to check that
$$
T_0 \frac{(k+\alpha_{0q},\alpha_{0q})
(k+\alpha_{nq},\alpha_{nq})}
{(k-\alpha_{0q},\alpha_{0q})
(k-\alpha_{nq},\alpha_{nq})} =
T_n \frac{(k+\alpha_{0q},\alpha_{0q})
(k+\alpha_{nq},\alpha_{nq})}
{(k-\alpha_{0q},\alpha_{0q})
(k-\alpha_{nq},\alpha_{nq})}
$$
at $\overline k_0 - \overline k_n +\overline e_0^2 - \overline e_n^2 =0$.
Equivalently we have
\begin{multline*}
\frac{(\overline k_0-\overline k_q+3\overline e_0^2+\overline e_q^2)(\overline k_n-\overline k_q+\overline e_n^2+\overline e_q^2)}{(\overline k_0-\overline k_q+\overline e_0^2-\overline e_q^2)
(\overline k_n-\overline k_q-\overline e_n^2-\overline e_q^2)}
=
\\=
\frac{(\overline k_0-\overline k_q+\overline e_0^2+\overline e_q^2)(\overline k_n-\overline k_q+3\overline e_n^2+\overline e_q^2)}
{(\overline k_0-\overline k_q-\overline e_0^2-\overline e_q^2)(\overline k_n-\overline k_q+\overline e_n^2-\overline e_q^2)}.
\end{multline*}
We express $\overline k_0$ through $\overline k_n$  and substitute the lengthes of
vectors. We obtain the correct equality
\begin{multline*}
\frac{(\overline k_n-\overline k_q-m-1)(\overline k_n-\overline k_q+m+1)}{(\overline k_n-\overline k_q+m-1)(\overline k_n-\overline k_q-m-1)}
=
\\=
\frac{(\overline k_n-\overline k_q+m+1)(\overline k_n-\overline k_q+3m+1)}{(\overline k_n-\overline k_q+3m+1)(\overline k_n-\overline k_q+m-1)}.
\end{multline*}

Finally, consider the last case

d) $1\le i\le n-1$, $j=n$. Like in the case b) we have to consider
the cases $q>0$ and $q=0$ separately. We assume at first that
$q>0$. We have to check that \beq{tin} (T_i-T_n)
\frac{(k+\alpha_{nq},\alpha_{nq})} {(k-\alpha_{nq},\alpha_{nq})}
\frac{\prod_{t=1}^m (k+t\alpha_{iq},\alpha_{iq})} {\prod_{t=1}^m
(k-t\alpha_{iq},\alpha_{iq})} =0 \eeq at $\overline k_i - \overline k_n +
\overline e_i^2 - \overline e_n^2 =0$. We consider separately $(T_i-T_n)$
applied to the numerator of (\ref{tin}). We get
\begin{multline*}
(\overline k_n-\overline k_q+m+1)\prod_{t=1}^m (\overline k_i - \overline k_q +2t+2) -
(\overline k_n-\overline k_q+3m+1)\prod_{t=1}^m (\overline k_i - \overline k_q +2t) =
\\=
\left((\overline k_i-\overline k_q+2m+2)(\overline k_i-\overline k_q+2) -
(\overline k_i-\overline k_q+2)(\overline k_i-\overline k_q+2m+2)\right)
\times
\\\times
\prod_{t=2}^m (\overline k_i-\overline k_q+2t)
= 0.
\end{multline*}
Analogously we get
$$
(T_i-T_n)(k-\alpha_{nq},\alpha_{nq})
\prod_{t=1}^m (k-t\alpha_{iq},\alpha_{iq}) =0,
$$
therefore condition (\ref{tin}) is satisfied. Finally let $q=0$. We have
to check that
$$
(T_i-T_n)
\frac{(k+\alpha_{i0},\alpha_{i0})(k+\alpha_{n0},\alpha_{n0})}
{(k-\alpha_{i0},\alpha_{i0})(k-\alpha_{n0},\alpha_{n0})} =0
$$
if $\overline k_i - \overline k_n +1-m=0$. We have
\begin{multline*}
(T_i-T_n)
\frac{(k+\alpha_{i0},\alpha_{i0})(k+\alpha_{n0},\alpha_{n0})}
{(k-\alpha_{i0},\alpha_{i0})(k-\alpha_{n0},\alpha_{n0})} =
\\[2mm]=
\frac{(\overline k_i-\overline k_0-m+2)(\overline k_n-\overline k_0-1)}{(\overline k_i-\overline k_0+m+2)(\overline k_n-\overline k_0+1)}
-
\frac{(\overline k_i-\overline k_0-m)(\overline k_n-\overline k_0+2m-1)}{(\overline k_i-\overline k_0+m)(\overline k_n-\overline k_0+2m+1)}
=
\\[2mm]=
\frac{\overline k_n-\overline k_0-1}{\overline k_i-\overline k_0+m+2} -
\frac{\overline k_i-\overline k_0-m}{\overline k_n-\overline k_0+2m+1} =0.
\end{multline*}
Lemma \ref{lm23} is fully proven.
\end{proof}

Now we are prepared for the proof of proposition \ref{predl7}.
\begin{proof}
At first we note that the operator $D$ can be represented in the form
$$
D=\sum_{p=0}^n \frac1{(\overline e_p, \overline e_p)}\prod_{\nad{q=0}{q\ne
p}}^n\vec\alpha_{qp} T_p \frac1{\prod_{\nad{q=0}{q\ne
p}}^n\vec\alpha_{qp}},
$$
where
$$
\vec\alpha_{qp} = \prod_{s=1}^{m_{qp}}
(k+s\alpha_{qp},\alpha_{qp}).
$$
We have to prove that
$$
(T^s_i-T^s_j)\frac{Df(k)}{\prod_{\nad{\beta\in
A_+}{\beta\ne\alpha_{ij}}}\vec\beta} = 0 \qquad  \mbox{   at  }\
k\in \pi_{ij},
$$
if
\beq{aksbl}
(T^s_i-T^s_j)\frac{f(k)}{\prod_{\nad{\b\in A_+}{\b\ne\a_{ij}}}\vec\beta} =0\qquad
\mbox{  at  }\
k\in\pi_{ij},
\eeq
$s=1,\ldots,m_{ij}.$

We have
$$
(T^s_i-T^s_j)\frac{Df(k)}{\prod\vec\beta} =
(T^s_i-T^s_j)\frac{1}{\prod\vec\beta}\sum_{p=0}^n\frac1{(\overline e_p, \overline e_p)}
\prod_{\nad{q=0}{q\ne
p}}^n\vec\alpha_{qp} T_p \frac1{\prod_{\nad{q=0}{q\ne
p}}^n\vec\alpha_{qp}}f(k).
$$
We show that the terms in the last sum corresponding to $p\ne i,j$
vanish, that is we show that \beq{aksbln}
(T^s_i-T^s_j)\frac{1}{\prod_{\nad{\beta\in
A_+}{\beta\ne\alpha_{ij}}}\vec\beta} \prod_{\nad{q=0}{q\ne
p}}^n\vec\alpha_{qp} T_p \frac1{\prod_{\nad{q=0}{q\ne
p}}^n\vec\alpha_{qp}}f(k)=0 \eeq at $k\in\pi_{ij}$. According to lemma
\ref{lm23} conditions (\ref{aksbln}) are equivalent for different
choices of ${A_+}$ such that $\alpha_{ij}$ is edge vector.
Therefore we can assume that  $A_+$ contains vectors $\alpha_{qp},
0\le q\le n,\, q\ne p$. For such a choice of $A_+$ one can carry
out cancellations in (\ref{aksbln}) and continue the equality
$$
(T^s_i-T^s_j)\frac{1}{\prod_{\nad{\beta\in A_+}{\beta\ne\alpha_{ij}, (\beta, e_p)=0}}\vec\beta}
 T_p \frac{f(k)}{\prod_{\nad{q=0}{q\ne
p}}^n\vec\alpha_{qp}}=
T_p(T_i^s-T_j^s)\frac{f(k)}{\prod_{\nad{\beta\in A_+}{\beta\ne\alpha_{ij}}}\vec\beta}
=0
$$
because of (\ref{aksbl}). Thus we get
\begin{multline}\label{aksblc}
(T^s_i-T^s_j)\frac{1}
{\prod_{\nad{\beta\in
A_+}{\beta\ne\alpha_{ij}}}} Df(k)
=\\=
(T^s_i-T^s_j)\frac1{\prod\vec\beta}
\bigg(\frac1{(\overline e_i,\overline e_i)}\prod_q \vec\alpha_{qi}T_i
\frac1{\prod_q\vec\alpha_{qi}}+
\frac1{(\overline e_j,\overline e_j)}\prod_q \vec\alpha_{qj}T_j
\frac1{\prod_q\vec\alpha_{qj}}\bigg)f(k).
\end{multline}
Because of lemma \ref{lm23} it is again  legal to prove the triviality of
the last expression for a special choice of $A_+$ only.
We choose  $A_+$ containing the vectors $\alpha_{qi},\,\alpha_{qj},\,0\le q\le n, q\ne i,j$. Then in equality (\ref{aksblc}) one can perform cancellations
and commutation such that the equality continues as follows
\begin{multline}\label{gmgm}
(T^s_i-T^s_j)\frac{1}{\prod_{\nad{\beta\in
A_+}{\beta\ne\alpha_{ij}}}} Df(k) =
\\=
(T^s_i-T^s_j)\Bigg(\frac{\vec \alpha_{ji}}{(\overline e_i,\overline e_i)}T_i
(\frac1{\vec\alpha_{ji}})T_i\frac1{\prod_{\nad{\beta\in
A_+}{\beta\ne\alpha_{ij}}}\vec\b} +
\frac{\vec \alpha_{ij}}{(\overline e_j,\overline e_j)}T_j
(\frac1{\vec\alpha_{ij}})T_j\frac1{\prod_{\nad{\beta\in
A_+}{\beta\ne\alpha_{ij}}}\vec\b}\Bigg) f(k)
=\\[2mm]
=\frac1{(\overline e_i,\overline e_i)}\frac{T^s_i\vec\alpha_{ji}}{T_i^{s+1}\vec\alpha_{ji}}
T_i^{s+1}\bigg(\frac{f(k)}{\prod\vec\beta}\bigg)-
\frac1{(\overline e_j,\overline e_j)}\frac{T^s_j\vec\alpha_{ij}}{T_j^{s+1}\vec\alpha_{ij}}
T_j^{s+1}\bigg(\frac{f(k)}{\prod\vec\beta}\bigg)
-\\[2mm]
-\frac1{(\overline e_i,\overline e_i)}\frac{T^s_j\vec\alpha_{ji}}{T^s_jT_i\vec\alpha_{ji}}
T_iT^s_j\bigg(\frac{f(k)}{\prod\vec\beta}\bigg) +
\frac1{(\overline e_j,\overline e_j)}\frac{T^s_i\vec\alpha_{ij}}{T^s_iT_j\vec\alpha_{ij}}
T_jT^s_i\bigg(\frac{f(k)}{\prod\vec\beta}\bigg).
\end{multline}
In order to check that the obtained expression is equal to zero we
analyze the possible cases. If $m_{ij}=1$ then $s=1$ and $T_i
\vec\alpha_{ji} = T_i(k+\alpha_{ji},\alpha_{ji})=\overline k_j-\overline
k_i+(\overline e_j,\overline e_j)-(\overline e_i,\overline e_i) =0$ if $k\in\pi_{ij}$.
Analogously  $T_i\vec\alpha_{ij}=0$, thus the first two terms in
(\ref{gmgm}) vanish. Also the last two terms in (\ref{gmgm})
cancel pairwise as
\begin{multline*}
\frac1{(\overline e_i,\overline
e_i)}\frac{T_j\vec\alpha_{ji}}{T_jT_i\vec\alpha_{ji}} =
\frac{T_j(k+\alpha_{ji},\alpha_{ji})}{(\overline e_i,\overline e_i)T_jT_i(k+\alpha_{ji},\alpha_{ji})}
= \frac{\overline k_j-\overline k_i+3e_j^2+e_i^2}{(\overline e_i,\overline e_i)(\overline k_j-\overline
k_i+3e_j^2-e_i^2)}=
\\=
\frac{2(e_i^2+e_j^2)}{2e_i^2e_j^2} = \frac1{(\overline e_j,\overline
e_j)}\frac{T_i\vec\alpha_{ij}}{T_iT_j\vec\alpha_{ij}}
\end{multline*}
at $k\in\pi_{ij}$. Now, if  $m_{ij}=m$, that is $1\le i,j \le n-1$, then
at $k_i=k_j$ because of symmetry obviously
$$
\frac1{(\overline e_i,\overline e_i)}\frac{T^s_i\vec\alpha_{ji}}{T^{s+1}_i\vec\alpha_{ji}}
=
\frac1{(\overline e_j,\overline
e_j)}\frac{T^s_j\vec\alpha_{ij}}{T^{s+1}_j\vec\alpha_{ij}}=g(k),
$$
and also
$$
\frac1{(\overline e_i,\overline e_i)}\frac{T^s_j\vec\alpha_{ji}}{T^{s}_jT_i\vec\alpha_{ji}}
=
\frac1{(\overline e_j,\overline e_j)}\frac{T^s_i\vec\alpha_{ij}}{T^{s}_iT_j\vec\alpha_{ij}}=h(k).
$$
Thus relation (\ref{gmgm}) can be rewritten as
\begin{multline*}
\left( T^s_i-T^s_j\right)\frac1{\prod\vec\beta} Df(k) =
g(k)\left(T_i^{s+1}\frac{f(k)}{\prod\vec\beta}-T_j^{s+1}\frac{f(k)}{\prod\vec\beta}\right)+
\\+
h(k)T_iT_j\left(T_i^{s-1}\frac{f(k)}{\prod\vec\beta} -
T_j^{s-1}\frac{f(k)}{\prod\vec\beta}\right) + O(k_i-k_j) = 0,
\end{multline*}
because conditions (\ref{aksbl}) hold, here we have $1\le
s<m_\alpha$. In the case $s=m_\alpha$ the previous equality also
takes place as in this case $g(k)=0$. The proposition is proven.
\end{proof}

\begin{theorem}
Let
\begin{multline}
\varphi_0=\left( (\overline k_0 - \overline k_n)^2 -1\right)\prod_{\nad{i,j=1}{i<j}}^{n-1} \prod_{s=1}^m \left( (k_i-k_j)^2 - 4 s^2 \right) \times
\\
\prod_{i=1}^{n-1} \bigl( (\overline k_0 - k_i)^2 - m^2 \bigr)
\bigl( (k_i - \overline k_n)^2 - (m+1)^2 \bigr),
\end{multline}
and let
$$
\varphi_{t+1}=(D-\lambda(x))\varphi_t,
$$
where
$D$ is operator $(\ref{DAnmC}),$
$\lambda(x)=\sum_{i=0}^n\frac1{\overline e_i^2} e^{2\overline x_i},$
and $t=0,1,2,\ldots$
Then
$$
\psi(k,x)= \bigg[2^M M! \prod_{i<j}(e^{2\overline x_i}-e^{2\overline
x_j})^{m_{ij}}\bigg]^{-1}\varphi_M(k,x)$$ is the Baker--Akhiezer
function for the configuration  $A_{n,2}(m)$ if
$M=\frac{m(n-1)(n-2)}2+2n-1$.
\end{theorem}

\begin{proof}
We note at first that function $\varphi_0$ has in fact the following form
$$\varphi_0
=\prod_{\alpha\in A_{n,2}(m)}\prod_{i=1}^{m_\alpha}(k+i\alpha,\alpha)(k-i\alpha,\alpha)e^{(k,x)}.
$$
Therefore propositions \ref{predl6}, \ref{predl7} guarantee  that for any $s$ function $\varphi_s(k,x)$ has the form
$\varphi_s=P_s(k,x)e^{(k,x)}$ where $P_s$ is a polynomial in $k$
with highest term $P_s^0$, and also $\varphi_s$ satisfies
axiomatics  (\ref{aksA0}), (\ref{aksA}). Thus we have to show that
if $s=M$ the first condition of the BA function definition
(\ref{psi12}) holds, that is $P_M^0 =\prod_{i<j} (\overline k_i-\overline
k_j)^{m_{ij}}$. For that we analyze how $P_s^0$ changes while one
applies operator $D-\lambda(x)$.

\begin{lemma}
Let $(D-\lambda(x))(Q_1(k,x)e^{(k,x)})=Q_2(k,x)e^{(k,x)},$ where
$Q_1,Q_2$ are polynomials in $k$ with the highest terms
$Q_1^0,Q_2^0$. Then \beq{lmst} Q_2^0 = 2\sum_{i=0}^n e^{2\overline
x_i}\frac{\partial Q_1^0}{\partial \overline k_i}+ \bigg(\sum_{i=0}^n
e^{2\overline x_i} \sum_{\nad{j=0}{j\ne i}}^n \frac{-2m_{ij}}{\overline
k_i-\overline k_j}\bigg)Q_1^0. \eeq
\end{lemma}
To prove the lemma we rewrite operator $D$ in the form
$$
D=\sum_{i=0}^n\frac1{\overline e_i^2}\prod_{\nad{j=0}{j\ne i}}^n
\frac{\overline k_i - \overline k_j -m_{ij}(\overline e_i^2+\overline e_j^2)}{\overline k_i-\overline
k_j+\overline e_i^2-\overline e_j^2}T_i.
$$
Now the arguments analogous to the ones given in the proof of theorem \ref{teorCsyst},
show that
\begin{multline*}
Q_2^0 =\sum_{i=0}^n\frac{e^{2\overline x_i}}{\overline e_i^2}\frac{\partial Q_1^0}{\partial k_i}
\,2\sqrt{(\overline e_i,\overline e_i)}+
\\+
\sum_{i=0}^n\frac{e^{2\overline
x_i}}{\overline e_i^2}\sum_{j\ne i} \frac1{\overline k_i - \overline k_j}
\left(-(m_{ij}+1)\overline e_i^2-(m_{ij}-1)\overline e_j^2\right)Q_1^0.
\end{multline*}
And it is easy to notice that the obtained expression coincides
with the one in formula~(\ref{lmst}). In particular, if $Q_1^0$ is
a linear combination of monomials,  $Q_1^0 =
\sum_{\{\lambda\}}\prod_{i<j} (\overline k_i-\overline k_j)^{\lambda_{ij}}$,
then \beq{q20} Q_2^0 = \sum_{\{\lambda\}}\sum_{i_0<j_0}
2(\lambda_{i_0j_0}-m_{i_0j_0})(e^{2\overline x_{i_0}}-e^{2\overline
x_{j_0}})(\overline k_{i_0}-\overline
k_{j_0})^{\lambda_{i_0j_0}-1}\prod_{(i,j)\ne(i_0,j_0)} (\overline
k_i-\overline k_j)^{\lambda_{ij}}. \eeq Thus in order to construct
$\varphi_i$ we start with monomial $P_0^0 =
\prod_{i<j}(k_i-k_j)^{2m_{ij}}$, and at every step $i$ the highest
term $P_i^0$ is a linear combination of monomials of the form
$\prod (\overline k_i-\overline k_j)^{\lambda_{ij}}$. From formula
(\ref{q20}) it can be seen that $\lambda_{ij}\ge m_{ij}$,
therefore at the step with the number $M=\sum m_\alpha$ it is
necessarily that $P_M^0 = C(x)\prod_{i<j}(\overline k_i -\overline
k_j)^{m_{ij}}$. Combinatorial arguments similar to the ones given
in \ref{teorCsyst} for the system $C_n(m,l)$ show that $C(x) =2^M
M! \prod_{i<j}(e^{2\overline x_i}-e^{2\overline x_j})^{m_{ij}}$, thus the
theorem is proven.
\end{proof}

In the end of this section we put the result on bispectrality.

\begin{theorem}
The Baker--Akhiezer function $\psi(k,x)$ for the system $ A_{n,2}(m)$ satisfies the following equation
in variables $k$:
$$
D \psi(k,x) = \sum_{i=0}^n\frac1{\overline e_i^2} e^{2\overline x_i} \psi (k,x),
$$
where $D$ is operator $(\ref{DAnmC})$.
For any polynomial $p(k)\in R_{A_{n,2}(m)}$
the difference operators
$$
D_p = ad_D^{\deg p} p(k)
$$
commute. These operators also commute with operator $D$.

\end{theorem}

\begin{proof}

By propositions \ref{predl6}, \ref{predl7} the function
$$
(D- \sum_{i=0}^n\frac1{\overline e_i^2} e^{2\overline x_i}) \psi(k,x)
$$
has the form $P(k,x)e^{(k,x)}$ where $P$ is a polynomial in $k$ of degree less than $\sum m_\a$, and satisfies axioms
(\ref{aksA0}), (\ref{aksA}). By lemma \ref{lem1} we have $P=0$ which is the
required  equation.
The proof of the second part of the theorem is also identical to the proof of
the corresponding theorem \ref{bispA2} about configuration $\cnm$.

\end{proof}

\section{Trigonometric locus conditions}
\label{gl2sec4}

In this section we obtain restrictions for a configuration ${\cal
A} = (A, m)$ to admit Baker--Akhiezer function. We obtain them
from the Scroedinger equation which holds for the BA function, the
restrictions turn out to be quite strong, they also have clear
geometrical sense.

By Proposition \ref{urS} we have the following equation for the Baker--Akhiezer
function $\psi(k,x)$:
\beq{psiur}
\left(\Delta-\sum_{\alpha\in A}
\frac{m_\alpha (m_\alpha+1) (\alpha,\alpha)}{\sinh^2 (\alpha,x)}
\right) \psi(k,x) = k^2 \psi(k,x).
\eeq
In paper \cite{CFV3} such equation was considered for arbitrary meromorphic
potential and function $\psi$ of the form $\psi = P(k,x) e^{(k,x)}$ where
$P$ is a polynomial in $k$. As it was shown in \cite{CFV3} (see also \cite{Phd}) potential
should satisfy the so called locus conditions. Regarding the form \mref{psiur}
these conditions have the form
\beq{triglocus}
\p_{\a}^{2s-1}\sum_{\nad{\b\in A}{\b\ne\a}}
\frac{m_{\b}(m_{\b}+1)(\b,\b)}{\sinh^2(\b,x)}=0 \qquad \mbox{at } \  \sinh(\a,x)=0,
\eeq
$s=1,\ldots,m_\a$. Â

We take vectors $\b$ forming  a positive subsystem $A_+$, with
$\a$ one of the edge vectors. Then projections
\beq{project}
\widehat\b=\b-a_\b\a, \quad a_\b=\frac{(\a,\b)}{(\a,\a)}
\eeq
of the
vectors $\b$ to the hyperplane $\Pi: (\a,x)=0$  belong to a
half-space in this hyperplane. Indeed, otherwise we have a
non-trivial dependence $\sum_{\b\in A_+\setminus \a} r_\b \widehat
\b =0 $ with some non-negative real coefficients $r_\b$. Then
$\sum_{\b\in A_+\setminus \a} r_\b  \b = \lambda \a $ for some
$\lambda\in\C$. Since all the vectors from $A_+$ belong to some
lattice it follows that coefficient $\lambda$ must be real. In
order for $\a$ to belong to the same half space as all $\b$ it
must be $\lambda>0$ which contradicts the condition that $\a$ is
an edge vector. So the projections $\widehat \b$ must belong to a
half-space. Denote by $\sigma$  the border of this half-space.

The cone $K=\{\mathrm{Re}(\widehat\b,x)<0|\ \b\in
{A_+}\backslash\a \}$ has non-empty intersection with $\Pi$.
Indeed, we consider generic extension to $\C^n$ of the
$(2n-3)-$plane $\sigma$ to form a $(2n-1)-$ hyperplane. Let it
have the equation  $Re (u,x) = 0$ for some $u\in \C^n$ so that $Re
(u,\widehat\beta) < 0$ for all projections $\widehat \b$. Now
consider $\widehat u = u  - \frac{(u,\a)}{(\a,\a)}\a$. One has
$\widehat u \in \Pi$, and also $(\widehat u, \widehat\beta)= (u,
\widehat\beta)$ thus $ \widehat u \in  K$ so the intersection
$K\cap \Pi$ is non-empty.

In the cone $K$ we can expand $\sinh (\beta,x)$ into the corresponding series
so that conditions \mref{triglocus} take the form
$$
\p_\a^{2s-1}\sum_{\nad{\b\in A_+}{\b\ne\a}}
4m_{\b}(m_{\b}+1)(\b,\b)\sum_{j=1}^\infty je^{2j(\b,x)}=0\qquad
\mbox{at} \  \sinh(\a,x)=0.
$$
More explicitly we obtain
\beq{yavno1}
\sum_{\nad{\b\in A_+}{\b\ne\a}}\sum_{j=1}^\infty
m_{\b}(m_{\b}+1)(\b,\b)(j(\alpha,\b))^{2s-1}je^{2j(\b,x)}=0
\eeq
at $\Pi_n: \{(\a,x)=\pi i n\}; \, n\in \Z$. We note that the intersection
$K\cap \Pi_n$ is non-empty for any $n$. Indeed, we represent
 $x$ in the form  $x=\frac{\pi i n}{(\a,\a)}\a+y$ for some vector $y$. The
 condition $x\in\Pi_n\cap K$ takes the form $(\a,y)=0$ and
$\mathrm{Re}(\b,x)<0$,  $\b\in A_+\backslash\a$.
In terms of  $y$ we get
$$
\mathrm{Re}(\b,x)=\mathrm{Re}\Big(\widehat\b+a_\beta\a,\ \frac{\pi i n}{(\a,\a)}\a+y\Big) =
\mathrm{Re}(\widehat\b,y)+\mathrm{Re}(a_\b\pi i n)<0.
$$
Thus $x$ belongs to $\Pi_n\cap K$ if and only if the corresponding $y=x-\frac{\pi i n}{(\a,\a)}\a$ satisfies $(\a,y)=0$ and also $\mathrm{Re}(\widehat\b,y)< {- \mathrm{Re}}(a_\b\pi i n)$. The last intersection is non-empty as it contains
a real multiple of any vector from the cone $\Pi\cap K$.

Now in the cone $\widehat K_n = \big\{y|\ (\a,y)=0,\
\mathrm{Re}(\widehat\b,y)< {- \mathrm{Re}}(a_\b\pi i n)\big\}$ the following
form of conditions (\ref{yavno1}) takes place:
\beq{usly}
\sum_{\nad{\b\in A_+}{\b\ne\a}}\sum_{j=1}^\infty
m_{\b}(m_{\b}+1)(\b,\b)
\, (j(\alpha,\b))^{2s-1}
\, j
\, e^{2ja_\b\pi i n}
\, e^{2j(\hat\b,y)}=0.
\eeq

We notice that vectors $\widehat\b$ belong to a lattice of rank $n-1$ in
the hyperplane $\Pi$. Indeed, considering if necessary a sub-lattice of the
original lattice in $\C^n$ we may assume that vector $\a$ is an integer multiple
of a basis vector of this lattice. The projections of other $n-1$ basis vectors
to $\Pi$ will generate a lattice in $\Pi$ containing vectors $\widehat\b$.

Let $e_1^*,\ldots, e_{n-1}^*$ be a basis of this lattice.
We claim that  coefficients at each particular exponent $(p_1 e_1^*+ \ldots
+ p_{n-1} e_{n-1}^*, x)$ in \mref{usly} after collecting the terms equal
zero.
Indeed, the cone $\widehat K_n$ contains a parallelepiped of the form
\beq{B_lambda}
B_\lambda = \big\{x|\  (x,e_j^*)\in [\lambda_j, \lambda_j+2\pi i
t_j],\
0\le t_j\le 1,\   j=1,\ldots,n-1 \big\}
\eeq
for some $\lambda_j\in \C$.
Multiplying the series \mref{usly} by $exp (-p_1 e_1^*- \ldots
- p_{n-1} e_{n-1}^*, x)$ and integrating it over $B_\lambda$ (which can be
done term by term as \mref{usly} is uniformly convergent on $B_\lambda$)
we conclude that all the terms are zero except the coefficient at the chosen
exponent, thus it should vanish as well.

Consider now the set of vectors $B^1 =
\{\b_1,\ldots,\b_p\}\subset A_+$ such that $\widehat \b_1 = \ldots =
\widehat \b_p$, and $j\widehat\b \ne \widehat \b_1$ for any $\b\in A_+$,
 $j\in \N$, $j>1$. Â By the previous argument  we conclude
$$
\sum_{\b\in B^1} m_\b(m_\b+1)(\b,\b)(\a,\b)^{2s-1}e^{2a_\b\pi i n}
= 0.
$$
Since  $n\in\Z$ is arbitrary the set $B^1$ is decomposed into subsets $B^1=B^1_1\cup\ldots\cup B^1_t$ such that $\forall
\b,\gamma\in B^1_l$ one hasÿ $e^{2a_\beta\pi i} =
e^{2a_\gamma\pi i}$ and
\beq{kras1}
\sum_{\b\in B^1_l} m_\b(m_\b+1)(\b,\b)(\a,\b)^{2s-1} = 0.
\eeq
We note that condition $e^{2a_\beta\pi i} =
e^{2a_\gamma\pi i}$ is equivalent to  $a_\b-a_\gamma =
n_{\b\gamma}\in\Z$. ÑUsing  $\widehat\b=\widehat\g$ and recalling \mref{project}
we get
\beq{kras2}
\b-\g=n_{\b\g}\a.
\eeq
Further, for any $j\ge 1$ we clearly have
$$
\sum_{\b\in B^1}
m_{\b}(m_{\b}+1)(\b,\b)
\, (j(\alpha,\b))^{2s-1}
\, j
\, e^{2ja_\b\pi i n}=0,
$$
and therefore identity (\ref{usly}) is valid with the summation over
$\b\in
\linebreak\in
A_+\backslash\a\backslash B^1$. ÒThus system
$A_+\backslash\a$ can be presented as a union of subsystems
$B^1\sqcup B^2\sqcup \ldots$ for each of which it is valid
(\ref{kras1}), (\ref{kras2}). ÌWe have proven the following

\begin{theorem}\lb{restr}
Let configuration ${\cal A} = (A, m)$ admit the Baker--Akhiezer
function. Let  $A_+\subset (A \cup (-A))$ be a positive subsystem
and let vector $\a\in A_+$ be an edge vector. Then the system of
vectors $A_+\backslash\a$ can be represented as a disjoint union
of {\rm ``}series{\rm "}\linebreak $A_+\backslash\a =
B_1\sqcup\ldots\sqcup B_N$ such that for all $l$, $1\le l\ \le N$
one has

$1)$
for anyÿ $\b,\g\in B_l$ the difference $\b-\g=n_{\b\g}\a,$ \ with
$n_{\b\g}\in \Z;$

$2)$ $\sum_{\b\in B_l} m_\b(m_\b+1)(\b,\b)(\a,\b)^{2s-1} =0,$ \ where
$1\le s \le m_\a$.

\noindent
These properties are equivalent to locus conditions \mref{triglocus}.
\end{theorem}


\begin{thebibliography}{37}




\bibitem{AMM}
Airault H., McKean H.P., Moser J.
{\em Rational and elliptic solutions of
the Korteweg-de Vries equation and a related many-body problem}
/\!/ Comm. Pure Appl. Math. 1977. V.~30. P.~95--178.


\bibitem{B}
Berest Yu.
{\em Huygens' principle and the bispectral problem}
/\!/  CRM Proceedings and Lecture Notes. 1998. V.~14. P.~11--30.


\bibitem{BC}
Burchnall J. L., Chaundy T. W.
{\em Commutative ordinary differential operators}
/\!/ Proc. London Math. Soc. 1922. V. 21. P. 420--440. Proc. Royal Soc. London.
1928. V. 118. P.557--583.




\bibitem{C}
Calogero F.
{\em Solution of the one-dimensional $n$-body problem with
quadratic and/or inversely quadratic pair potential}
/\!/ J. Math.Phys. 1971. V.~12. P.~419--436.


\bibitem{Ch}
Chalykh O.A. {\em Darboux transformations for multidimensional
Schr\"odinger operators}  /\!/ Russian Math. Surveys. 1998. V. 53.
N. 2. P. 167--168.


\bibitem{ChalBisp}
Chalykh O.
{\em Bispectrality for quantum Ruijsenaars model and its integrable
deformations}
/\!/  J. Math. Phys. 2000. V. 41. N. 8. P. 5139--5167.




\bibitem{Ch3}
Chalykh O.A.
{\em Macdonald polynomials and algebraic integrability}
/\!/ Adv. Math. 2002. V. 166. N. 2. P. 193--259.


\bibitem{CEO}
Chalykh O., Etingof P., Oblomkov A.
{\em Generalized Lame operators}
/\!/ Commun. Math. Phys. 2003. V.~239. N. 1-2.  P.~115--153.


\bibitem{CFV2}
Chalykh O.A., Feigin M.V., Veselov A.P.
{\em New integrable generalisations of  Calogero--Moser quantum
problem}
/\!/ J. of Math. Phys. 1998. V.~39. N.~2. P.~695--703.


\bibitem{CFV3}
Chalykh O.A., Feigin M.V., Veselov A.P.
{\em Multidimensional Baker--Akhiezer Functions and Huygens' Principle}
/\!/  Commun. Math. Phys. 1999. V.~206.
P.~533--566.


\bibitem{ChV1}
Chalykh O.A., Veselov A.P.
{\em Commutative rings of partial differential operators and Lie
algebras}
/\!/ Commun. Math. Phys. 1990. V.~126. P.~597--611.


\bibitem{CV2}
Chalykh O.A., Veselov A.P.
{\em Locus configurations and $\vee$-systems}
/\!/ Phys. Let. A.  2001. V. 285  N. 5-6.  P. 339--349.


\bibitem{Cher}
Cherednik I.,
{\em Macdonald's evaluation conjectures and difference Fourier transform}
/\!/ Invent. Math. 1995. V. 122. P. 119--145.


\bibitem{DG}
Duistermaat J.J., Gr\"unbaum F.A.
{\em Differential equations with the spectral parameter}
/\!/ Comm. Math. Phys. 1986. V.~103. P.~177--240.

\bibitem{Phd}
Feigin M.
{\em Multidimensional integrable Schroedinger operators}
/\!/ 2001, Candidate of Science (PhD) thesis, Moscow State University.

\bibitem{H}
Hietarinta J.
{\em Pure quantum integrability}
/\!/ Phys. Lett. A  1998. V.246, P.~97--104.


\bibitem{Koorn1}
Koornwinder T. H., unpublished manuscript, 1988.


\bibitem{Koorn2}
Koornwinder T. H.
{\em Askey--Wilson polynomials for root systems of type $BC$}
/\!/ Contemp. Math. 1992. V.138. P. 189--204.



\bibitem{K}
Krichever I.M.
{\em  Methods of algebraic geometry in the theory of non-linear equations}
/\!/ Russian Math. Surveys  1977. ÒV.~32, N.~6. ÑP.~183--208.



\bibitem{Mac}
Macdonald I. G.
{\em Symmetric Functions and Hall Polynomials}
/\!/ 1995. Clarendon, Oxford, 2nd ed.

\bibitem{Macdonald}
Macdonald I.G.
{\em Orthogonal polynomials associated with root systems}
/\!/ 1988. Preprint.


\bibitem{M}
Moser J.
{\em Three integrable hamiltonian systems connected with
isospectral  deformations}
/\!/ Adv. Math. 1975. V. 16, P. 197--220.



\bibitem{OP}
Olshanetsky M.A., Perelomov A.M.
{\em Quantum integrable systems related to Lie algebras}
/\!/ Phys. Rep. 1983. V.~94. P.~313--404.




\bibitem{R1}
 Ruijsenaars S. N. M.
{\em Complete integrability of relativistic Calogero--Moser systems and elliptic functions identities}
/\!/ Commun. Math. Phys. 1987. V. 110. P. 191--213.


\bibitem{R2}
Ruijsenaars S. N. M.
{\em Finite-dimensional soliton systems," in Integrable and Superintegrable Systems}
/\!/ 1990. World Scientific, Singapore, edited by B. Kupershmidt,   P. 165--206.



\bibitem{RS}
Ruijsenaars S. N. M., Sneider H.
{\em A new class of integrable systems and its relation to solitons}
/\!/ Ann. Phys. 1986. V. 170. P. 370--405.

\bibitem{Sahi}
Sahi S.
{\em Nonsymmetric Koornwinder polynomials and duality}
/\!/ Ann. Math. 1999. V. 150.  P. 267--282.


\bibitem{S}
Sutherland B.
{\em Exact results for a quantum many-body problem in one
dimension}
/\!/ Phys. Rev. A. 1971. V.4.  P. 2019--2021.

\bibitem{vD}
van Diejen J. F.
{\em Self-dual Koornwinder-Macdonald polynomials}
/\!/ Invent. Math. 1996. V. 126. N.2. P. 319--339.


\bibitem{CFV1}
Veselov A.P., Feigin M.V., Chalykh O.A.
{\em New integrable deformations of quantum Calogero--Moser problem}
/\!/ Russian Math Surveys  1996. V.~51. N.3. P.~185--186.




\bibitem{VSCh}
 Veselov A.P., Styrkas K.L., Chalykh O.A.
 {\em Algebraic integrability for Schrodinger equation and finite reflection
k groups}
 /\!/ Theor. Math. Phys. 1993. V.~94, N.~2. P.~253--275.





\end{thebibliography}
\end{document}